\documentclass[epj,final]{svjour}

\usepackage{graphicx}
\usepackage{amsmath}
\usepackage{amssymb}
\usepackage{xcolor}

\begin{document}

\title{Spatio-temporal dynamics of an active, polar, viscoelastic ring}
\author{Philippe Marcq}                  

\institute{Physico-Chimie Curie, 
Institut Curie, CNRS, Universit\'e Pierre et Marie Curie,
26 rue d'Ulm, F-75248 Paris Cedex 05, France
\email{philippe.marcq@curie.fr}
}
%
\date{February 17, 2014}
%
\abstract{
Constitutive equations  for a one-dimensional, active, polar, 
viscoelastic liquid are derived by treating the 
strain field as a slow hydrodynamic variable.
Taking into account the couplings between strain and polarity 
allowed by symmetry, the hydrodynamics of an active, polar, 
viscoelastic body include an evolution equation for the polarity field 
that generalizes the damped Kuramoto-Sivashinsky equation. 
Beyond thresholds of the active coupling coefficients 
between the polarity and the stress or the strain rate, 
bifurcations of the homogeneous state lead first to stationary waves, 
then to propagating waves of the strain, stress and polarity fields.
I argue that these results are relevant to living matter,
and may explain rotating actomyosin rings in cells
and mechanical waves in epithelial cell monolayers.
\\
}

\maketitle

\section{Introduction}
\label{sec:intro}

At molecular scales, the mechanics and self-organization of the cytoskeleton
are driven by a chemical fuel, through ATP or GTP hydrolysis, 
and rely on the enzymatic activity of molecular motors to turn 
chemical energy into forces and motion.
Numerous phenomena pertaining to cell division, cell motility,
cell adhesion, or more generally pattern formation at the scale 
of the cytoskeleton, 
have been interpreted within the framework of hydrodynamics, where 
continuous fields subsume the behaviour of many individual active units, 
and governing equations are constrained by 
symmetry \cite{Juelicher2007,Ramaswamy2010,Marchetti2013}. 
Since biofilaments such as filamentous actin and microtubules 
are polar, orientational order must be considered. 
Cell rheology is characterized by a weak
power-law frequency dependence of the response functions 
and by nonlinear behaviour, such as strain-stiffening \cite{Chen2010}.
These complex features are difficult to treat analytically, rendering
simpler rheologies all the more attractive to the theorist.
The behaviour of active polar liquids has been studied in great 
detail \cite{Kruse2004,Kruse2005,Muhuri2007,Giomi2008}, 
while recent work has also dealt with the mechanical and dynamical properties
of isotropic active solids \cite{Banerjee2011,Edwards2011,Yoshinaga2012},
and of active polar permeating gels  \cite{Callan-Jones2011}.

Motivated by the observation of graded and alternating
polarity profiles of filamentous actin in the actomyosin bundles
of metazoan cells \cite{Yoshinaga2010}, we studied  pattern formation 
in an active polar solid band, taking into account possible
couplings between strain and polarity. We showed 
that the dynamics of the polarity field is
governed by a damped Kuramoto-Sivashinsky equation \cite{Manneville1988}.
For large damping, we found that an activity-driven 
stationary bifurcation leads to periodic patterns, suggesting a scenario
for the emergence of sarcomeric organization in actomyosin bundles.

Here I treat the case of an active, polar, viscoelastic (Maxwell) liquid 
in one spatial dimension, considering the strain field as a
hydrodynamic variable \cite{Brand1990}. 
For simplicity, the dynamical equations
are studied analytically on an infinite line,
and numerically with periodic boundary conditions:
this study pertains to dynamical behaviour on a ring, or far from the
boundaries in a quasi-one-dimensional system.
One objective is to elaborate on the relevance 
of the damped Kuramoto-Sivashinsky equation to the dynamics 
of active polar materials. 
A study of the complete parameter space is beyond the scope of this work,
where I show representative examples of the dynamics.

This article is organized as follows.
In section~\ref{sec:const}, I derive the set of coupled evolution 
equations for the velocity, strain and polarity fields that 
govern the dynamics of  an active polar viscoelastic liquid
in one spatial dimension.
In section~\ref{sec:stab}, I perform the linear stability analysis of the
homogeneous state, uncovering an activity-driven stationary bifurcation, then
show numerically that propagating waves also arise 
farther from the bifurcation threshold.
As summarized in the Appendix, I performed the same analysis in 
the case of an active, polar, viscoelastic (Kelvin) solid, and obtained
similar results. In section~\ref{sec:appl}, I discuss applications 
to living matter, at the scale of the cell
with actomyosin bundles and rings, 
but also at the scale of the tissue with epithelial cell monolayers.

\onecolumn

\section{Constitutive equations}
\label{sec:const}

In this section, I derive constitutive equations for an
active, polar, viscoelastic liquid in one spatial dimension.
The relevant fields include the pressure $P$, the temperature $T$, 
the total stress $\sigma$, the strain $e$, the velocity $v$
and the polarity $p$, all dependent upon the spatial and 
temporal coordinates $(x,t)$. 
I derive constitutive equations 
expressing first the material's free energy density as a Landau expansion
around the zero strain, zero polarity homogeneous state,
second the entropy production rate density as a function of the relevant
thermodynamic force-flux pairs, and finally fluxes as a function 
of forces, including coupling terms allowed by symmetry \cite{Chaikin1995}.

The free energy density, expanded 
to quadratic order about $p = e = 0$,  reads
\begin{equation}
  \label{eq:freeenergy}
  f = \frac{a}{2} \; p^2 + 
      \frac{K}{2} \; \left( \partial_x p \right)^2 +
      \frac{G}{2} \; e^2+
      w \; e \; \partial_x p
\end{equation}
where all terms are invariant under simultaneous inversion of space 
and polarity,  and the polar term 
$\partial_x p$ is  omitted since its contribution is equal to zero for 
periodic boundary conditions.
The positive coefficients $a$, $K$, and $G$
respectively denote the susceptibility, 
the energy cost of inhomogeneities of the polarity field,
and the elastic modulus. Symmetries of the problem 
allow for a coupling term between strain and polarity gradient,
with a coupling parameter $w$ of arbitrary sign.
Thermodynamic stability imposes the constraint $w^2 \le K  G$.
Differentiating eq.~(\ref{eq:freeenergy}) yields the molecular field $h$,
conjugate to the polarity
\begin{equation}
  \label{eq:def:h}
h = -\frac{\delta f}{\delta p} = - \; a \; p 
+ K \; \partial^2_x p +w \; \partial_x e
  \end{equation}
and the elastic stress, conjugate to the strain
\begin{equation}
  \label{eq:def:sigma:el}
\sigma^{\mathrm{el}} = \frac{\delta f}{\delta e} = G \; e + w \; \partial_x p.  
\end{equation}

As was first advocated in \cite{Brand1990} in the context of the viscoelastic
relaxation of the transient (but long-lived) strain of polymer melts,
I treat the strain field as an additional slow, hydrodynamic variable. The
same approach was used recently to derive in a natural and rigorous manner 
the constitutive equations of active polar 
permeating gels  \cite{Callan-Jones2011}.
The density $R/T$ of the entropy production rate is obtained by
standard manipulations of the conservation equations:
\begin{equation}
  \label{eq:R}
  R = \left( \sigma + P \right) \; \partial_x v 
      - \dot{e} \; \sigma^{\mathrm{el}} 
      + \dot{p} \; h + r \; \Delta \mu.
\end{equation}
This expression contains two mechanical terms, a polar term, 
and a (chemical) active term where the generalized force $\Delta \mu$ 
(variation of chemical potential due to ATP hydrolysis)  
is conjugate to the ATP consumption rate $r$.
A dot denotes the total derivative, as in
$\dot{p} = \partial_t p + v \, \partial_x p $.
The conjugate flux-force pairs are
\begin{eqnarray} 
\nonumber
 \mathrm{Flux} &\leftrightarrow& \mathrm{Force} \\
\nonumber
\sigma + P  &\leftrightarrow& \partial_x v \\
\label{eq:fluxforce}
\dot{e} &\leftrightarrow& \sigma^{\mathrm{el}} \\
\nonumber
\dot{p} &\leftrightarrow& h \\
\nonumber
r &\leftrightarrow& \Delta \mu.
\end{eqnarray}
Following a standard procedure \cite{Juelicher2007,Kruse2005,Callan-Jones2011},
the constitutive equations read
\begin{eqnarray}
  \label{eq:consteqsig}
  \sigma  +  P  &=&  \eta \; \partial_x v +  \sigma^{\mathrm{el}}  
           + \sigma_{\mathrm{a}} + \beta_{\mathrm{a}} \; \partial_x p \\
  \label{eq:consteqe}
  \dot{e}  &=& \partial_x v - \Gamma_e \; \sigma^{\mathrm{el}} 
           + {\dot e}_{\mathrm{a}} - \psi_{\mathrm{a}} \; \partial_x p \\
  \label{eq:consteqp}
  \dot{p} &=& \Gamma_{\mathrm{p}} \; h - \alpha_{\mathrm{a}}  \; p \, \partial_x p
- \lambda  \; p \, \partial_x v 
\end{eqnarray}
omitting the equation expressing as a function of thermodynamic forces
the field $r$.
The dynamic viscosity $\eta$ and the kinetic coefficients 
$\Gamma_e$ and $\Gamma_{\mathrm{p}}$ are positive.
Active coupling coefficients are denoted by the index $a$, and are
proportional to $\Delta \mu$, as in $\beta_{\mathrm{a}} = \beta \, \Delta \mu$.
The active stress $\sigma_{\mathrm{a}}$ is positive for a contractile material.
The invariance properties
of polar media allow for an active coupling between 
stress and polarity gradient, with a coefficient
$\beta_{\mathrm{a}}$, as first introduced in \cite{Giomi2008}.
The active coefficients ${\dot e}_{\mathrm{a}}$ and $\psi_{\mathrm{a}}$ in 
eq.~(\ref{eq:consteqe}) play the same role as $\sigma_{\mathrm{a}}$ 
and $\beta_{\mathrm{a}}$ in eq.~(\ref{eq:consteqsig}).
Two advection terms with coupling coefficients 
$\alpha_{\mathrm{a}}$ and $\lambda$, of arbitrary signs,
are included in eq.~(\ref{eq:consteqp}), in agreement with
earlier studies of active liquid crystals \cite{Muhuri2007,Giomi2008}.
Except for $\sigma_{\mathrm{a}}$ and ${\dot e}_{\mathrm{a}}$, the active terms
in (\ref{eq:consteqsig}-\ref{eq:consteqp}) contain products of $\Delta \mu$ 
with $p$ and its gradients, and are therefore \emph{stricto sensu}
nonlinear terms. In the long time limit, a non-zero value of 
${\dot e}_{\mathrm{a}}$ would lead to unbounded values of the strain,
unphysical in the regime of small deformations 
considered here.
In the following, I therefore set ${\dot e}_{\mathrm{a}} = 0$.

Three relaxation times characterize this material:
the viscoelastic time $\tau = \eta/G$, 
the strain relaxation time $\tau_{\mathrm{e}} = (G \Gamma_e)^{-1}$, 
and the polarity relaxation time 
$\tau_{\mathrm{p}} = (a \Gamma_{\mathrm{p}})^{-1}$.
A separation of time scales may lead to distinct dynamical regimes 
since the viscoelastic time, the strain relaxation time and the polarity
relaxation time may differ. 
Assuming a constant pressure and a constant active stress,
the momentum conservation equation  $\partial_x \sigma = 0$ yields
$\tau \; \partial^2_x v + \partial_x e + l_{\mathrm{p}} \; \partial^2_x p = 0$,
where $l_{\mathrm{p}} = \frac{w + \beta_{\mathrm{a}}}{G}$ is a length scale 
that characterizes the coupling between strain and polarity gradient.
Anticipating the possibility of an instability of the homogeneous state 
$v = e = p = 0$ at short wavelengths \cite{Yoshinaga2010}, I introduce
a stabilizing higher-order polarity gradient term 
$\frac{\nu}{2} \, \left( \partial^2_x p \right)^2$ in the free energy  
density, where $\nu$ is a positive coefficient \cite{Chaikin1995}.
As a consequence, the conjugate field $h$, eq.~(\ref{eq:def:h}), 
is supplemented with the term $- \nu \;  \partial^4_x p$.
Altogether, the dynamics is governed by a set of three coupled partial
differential equations  
\begin{eqnarray}
  \label{eq:dynsig}
  0  &=&   \tau \; \partial^2_x v + \partial_x e +
 l_{\mathrm{p}} \; \partial^2_x p \\
  \label{eq:dyne}
\partial_t e + v \, \partial_x e   &=& -\frac{1}{\tau_{\mathrm{e}}} e
+ \partial_x v   - v_{\mathrm{p}} \, \partial_x p \\
  \label{eq:dynp}
\partial_t p + v \, \partial_x p 
+ \lambda  \; p \, \partial_x v 
+ \alpha_{\mathrm{a}}  \; p \, \partial_x p
   &=& 
-\frac{1}{\tau_{\mathrm{p}}} p 
+ \Gamma_{\mathrm{p}} \left( K \, \partial_x^2 p 
- \nu  \, \partial^4_x p + w \, \partial_x e \right)
\end{eqnarray}
where $v_{\mathrm{p}} = \psi_{\mathrm{a}} + \Gamma_e w$ is a 
coefficient of arbitrary sign which characterizes the coupling between 
strain rate and polarity.

Given the definition (\ref{eq:def:sigma:el}) of the elastic stress
$\sigma_{\mathrm{el}}$, its evolution equation is obtained by combining
Eqs.~(\ref{eq:dyne}) and (\ref{eq:dynp}):
\begin{equation}
  \label{eq:dynsigmael}
{\dot \sigma_{\mathrm{el}}} + \frac{\sigma_{\mathrm{el}} }{\tau_{\mathrm{e}}} = 
 G \; \partial_x v 
+ \left( \frac{w}{\tau_{\mathrm{e}}} - G v_{\mathrm{p}} \right)\, \partial_x p 
+ w \, \partial_x {\dot p}.
\end{equation}
When the time $t$ is long compared to $\tau_{\mathrm{e}}$ and $\tau_{\mathrm{p}}$,
${\dot \sigma_{\mathrm{el}}}$ and ${\dot p}$ may relax to zero.
In this limit, the elastic stress becomes a linear combination of the velocity 
and polarity gradients
$\sigma_{\mathrm{el}}^{\infty} \to  G \tau_{\mathrm{e}} \; \partial_x v^{\infty} 
+ \left( w - G \tau_{\mathrm{e}} v_{\mathrm{p}} \right)\, 
\partial_x p^{\infty}$, and a stationary solution $\sigma^{\infty}(x)$
for the total stress reads (see eq.~(\ref{eq:consteqsig}):
\begin{equation}
  \label{eq:stat:stress}
  \sigma^{\infty}(x) =-  P +  
\left( \eta + G \tau_{\mathrm{e}} \right) \partial_x v^{\infty} 
           + \sigma_{\mathrm{a}} + 
\left( \beta_{\mathrm{a}} +w - G \tau_{\mathrm{e}} v_{\mathrm{p}} \right) 
\partial_x p^{\infty}.
\end{equation}
Assuming that this stationary solution  is stable,
the constitutive equation for the stress field (\ref{eq:stat:stress}) 
is that of an  active, polar liquid in the long time limit,
as expected for a Maxwell viscoelastic liquid.

\begin{figure}[!t]
\includegraphics*[width=0.43\columnwidth]{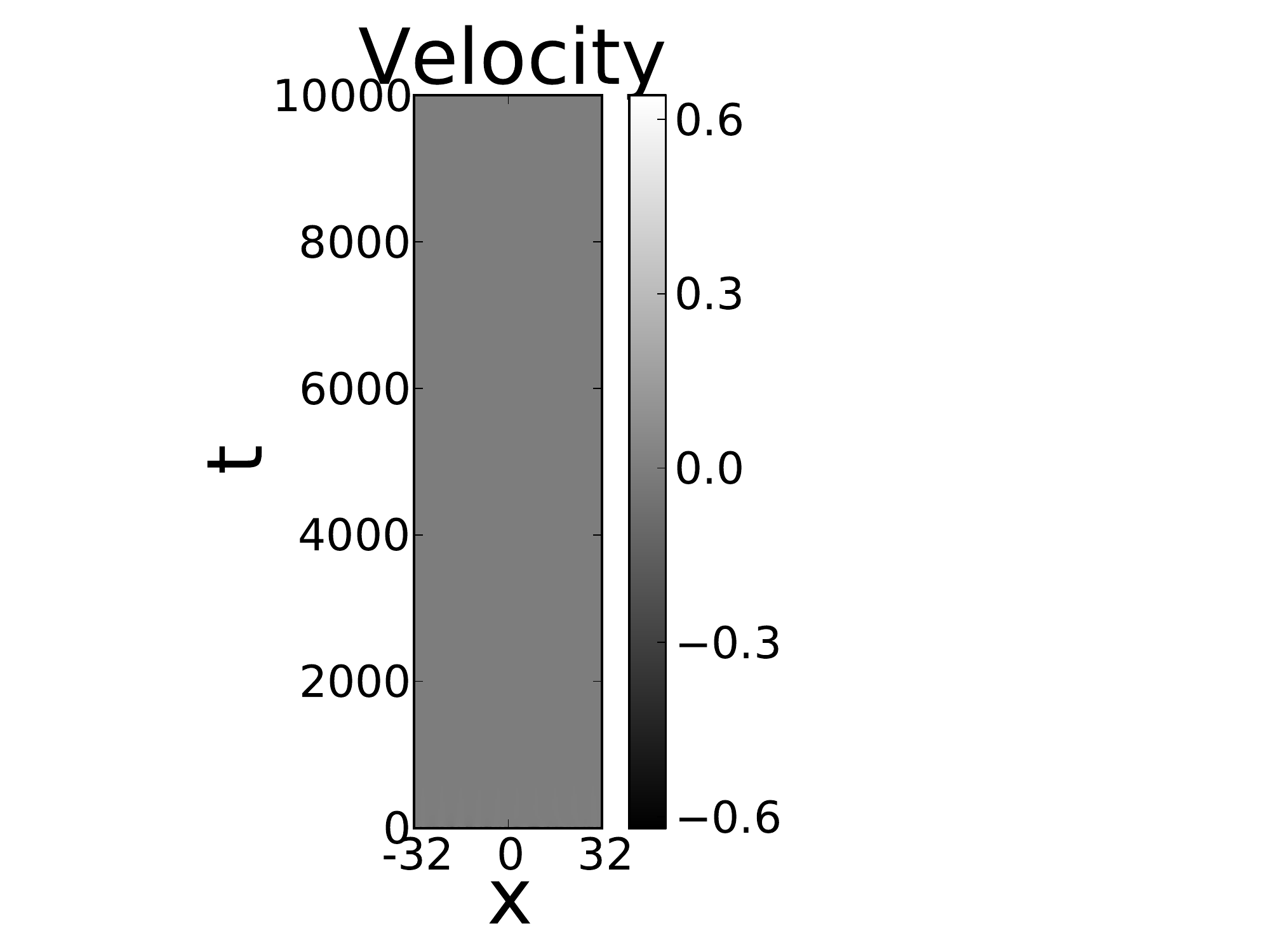}
\hspace*{-3.cm}
\includegraphics*[width=0.43\columnwidth]{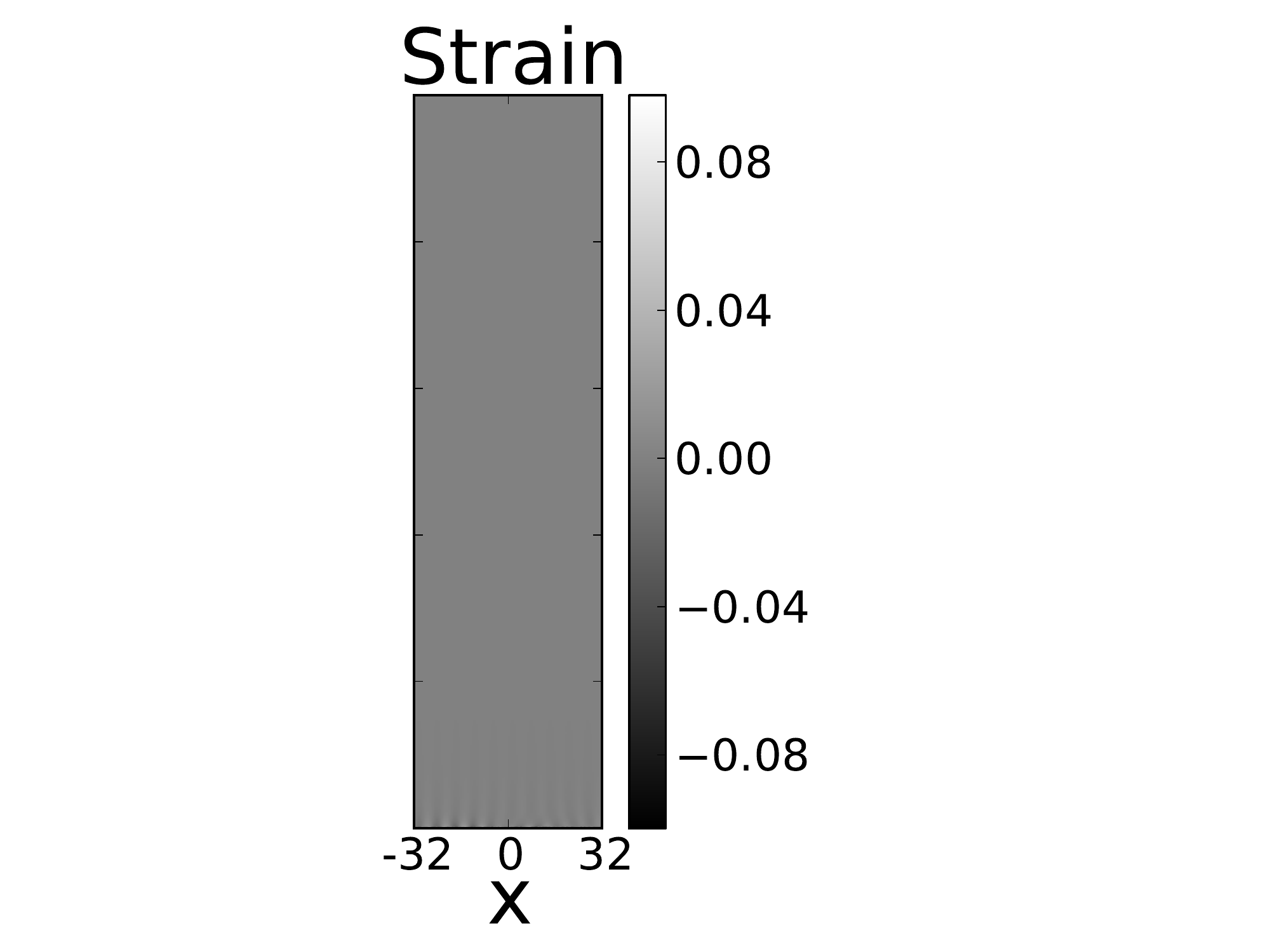}
\hspace*{-3.cm}
\includegraphics*[width=0.43\columnwidth]{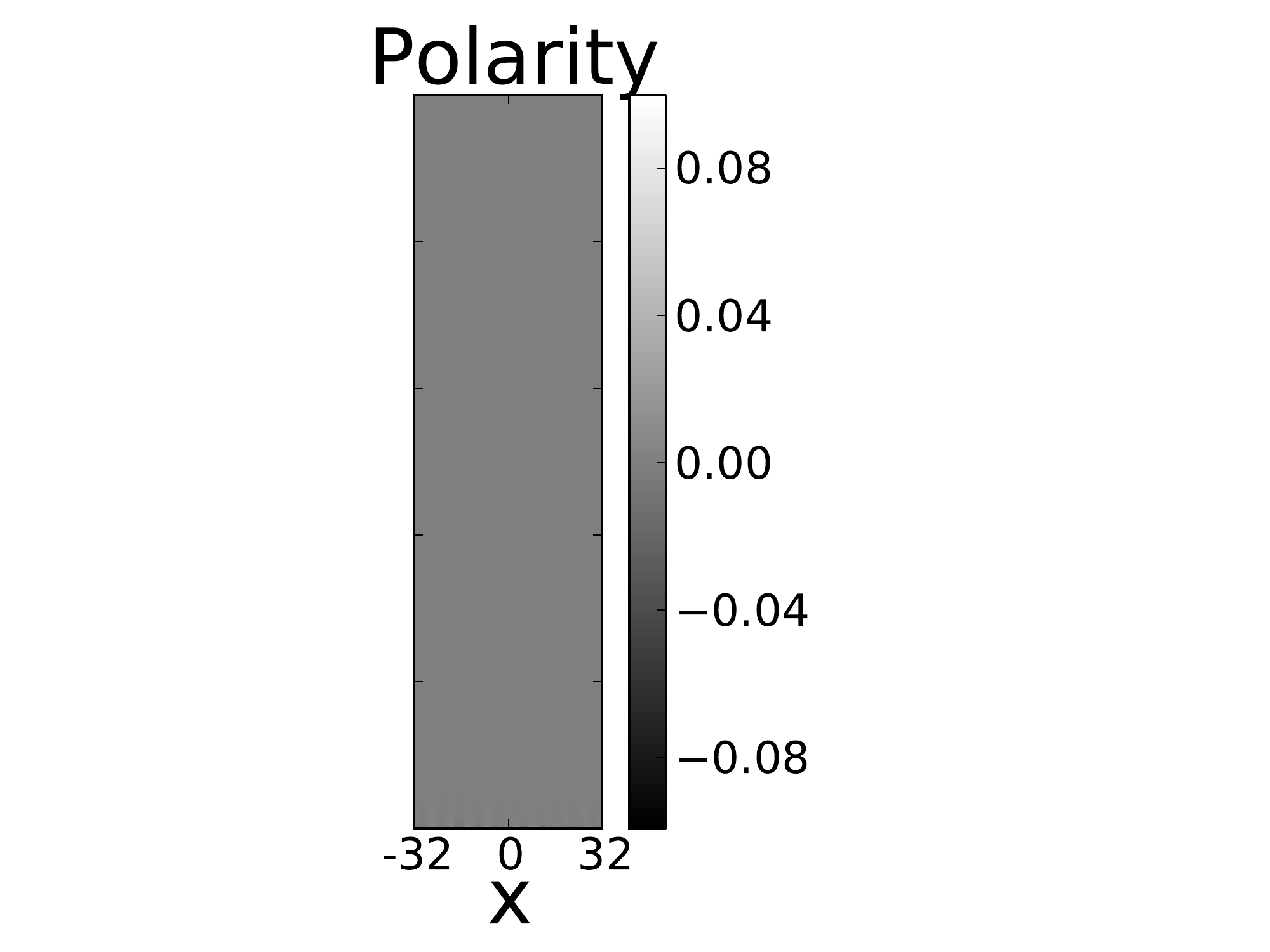}
\hspace*{-3.cm}\textbf{(a)}\\
\includegraphics*[width=0.43\columnwidth]{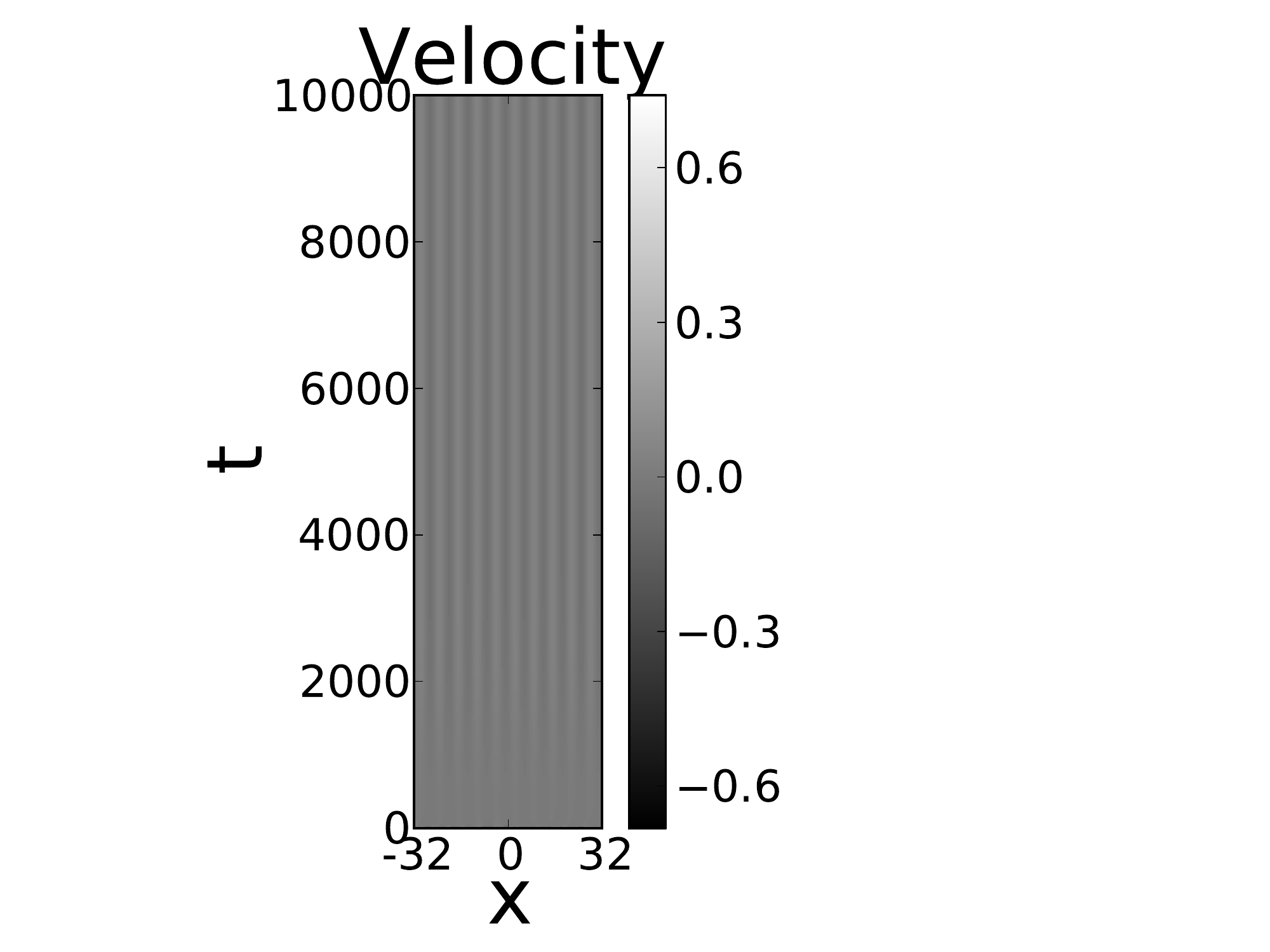}
\hspace*{-3.cm}
\includegraphics*[width=0.43\columnwidth]{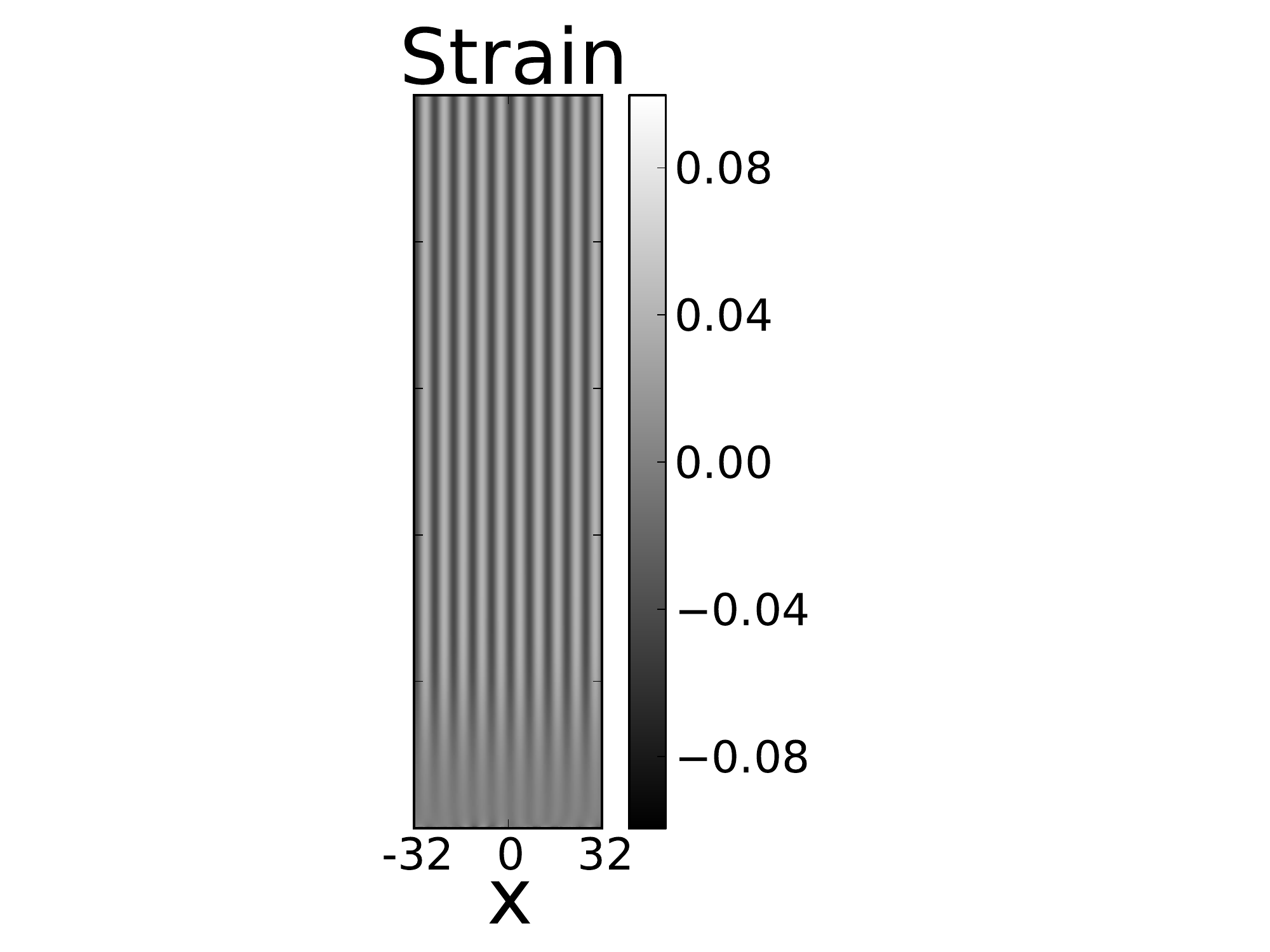}
\hspace*{-3.cm}
\includegraphics*[width=0.43\columnwidth]{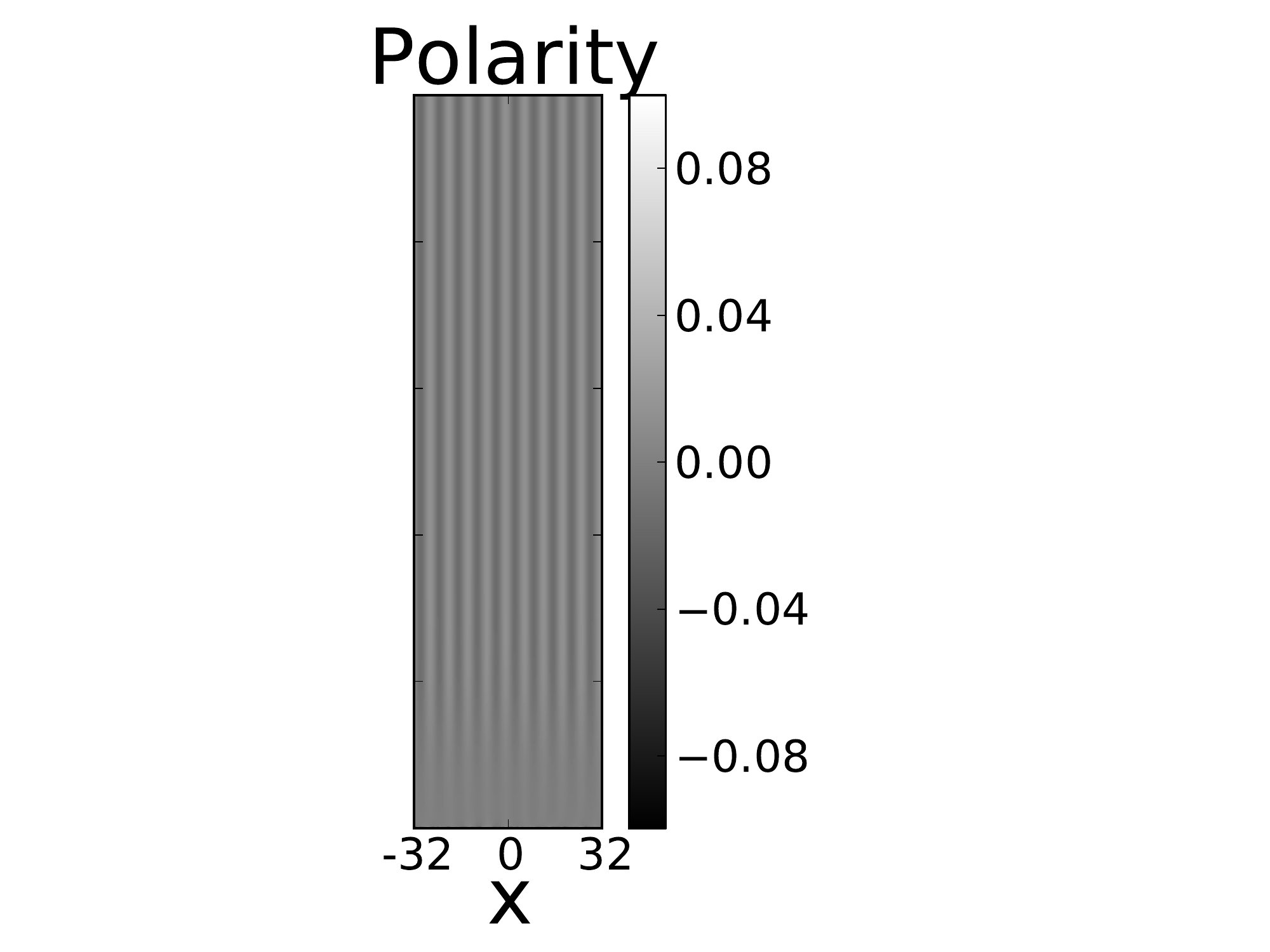}
\hspace*{-3.cm}\textbf{(b)}\\
\includegraphics*[width=0.43\columnwidth]{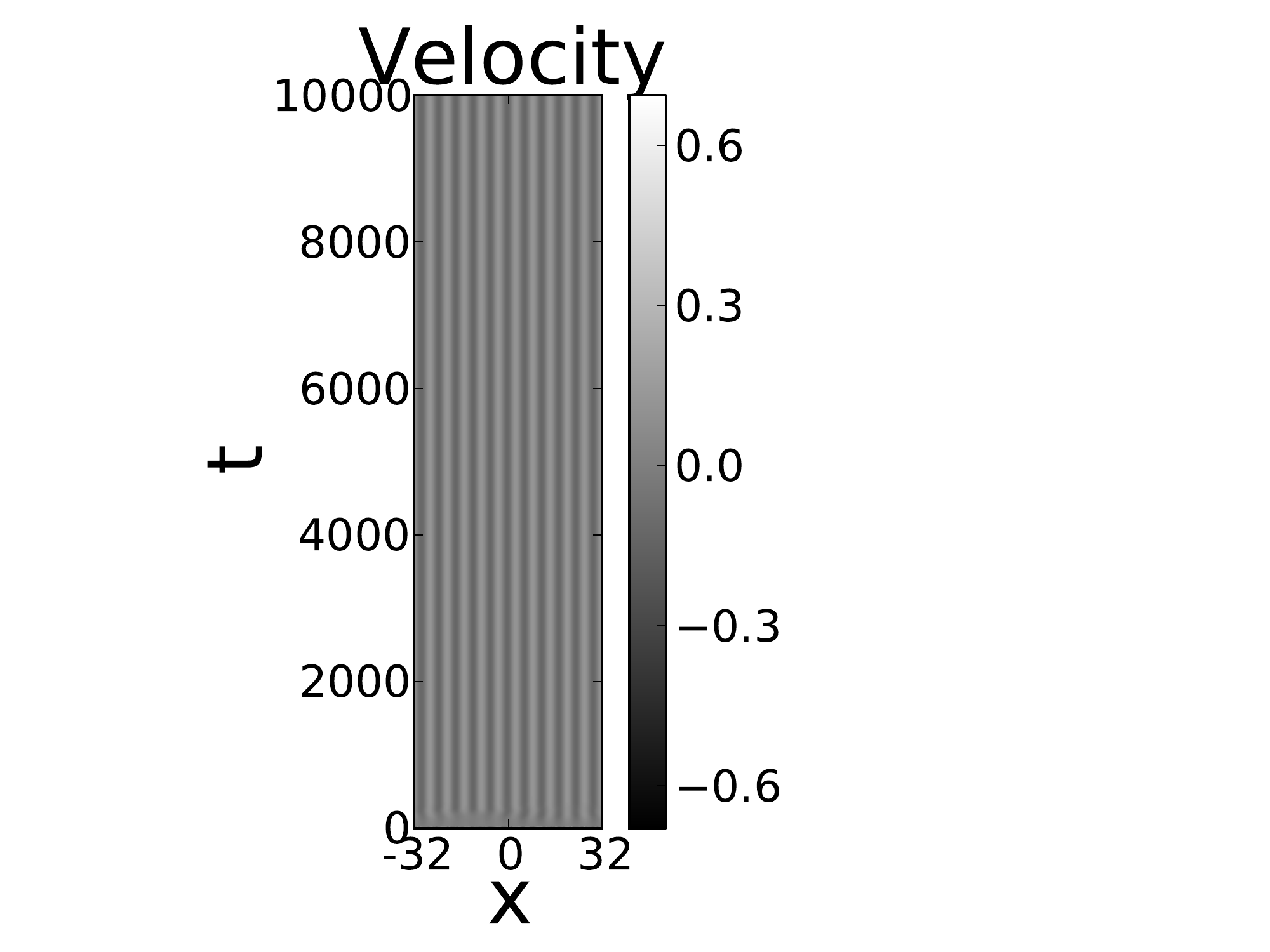}
\hspace*{-3.cm}
\includegraphics*[width=0.43\columnwidth]{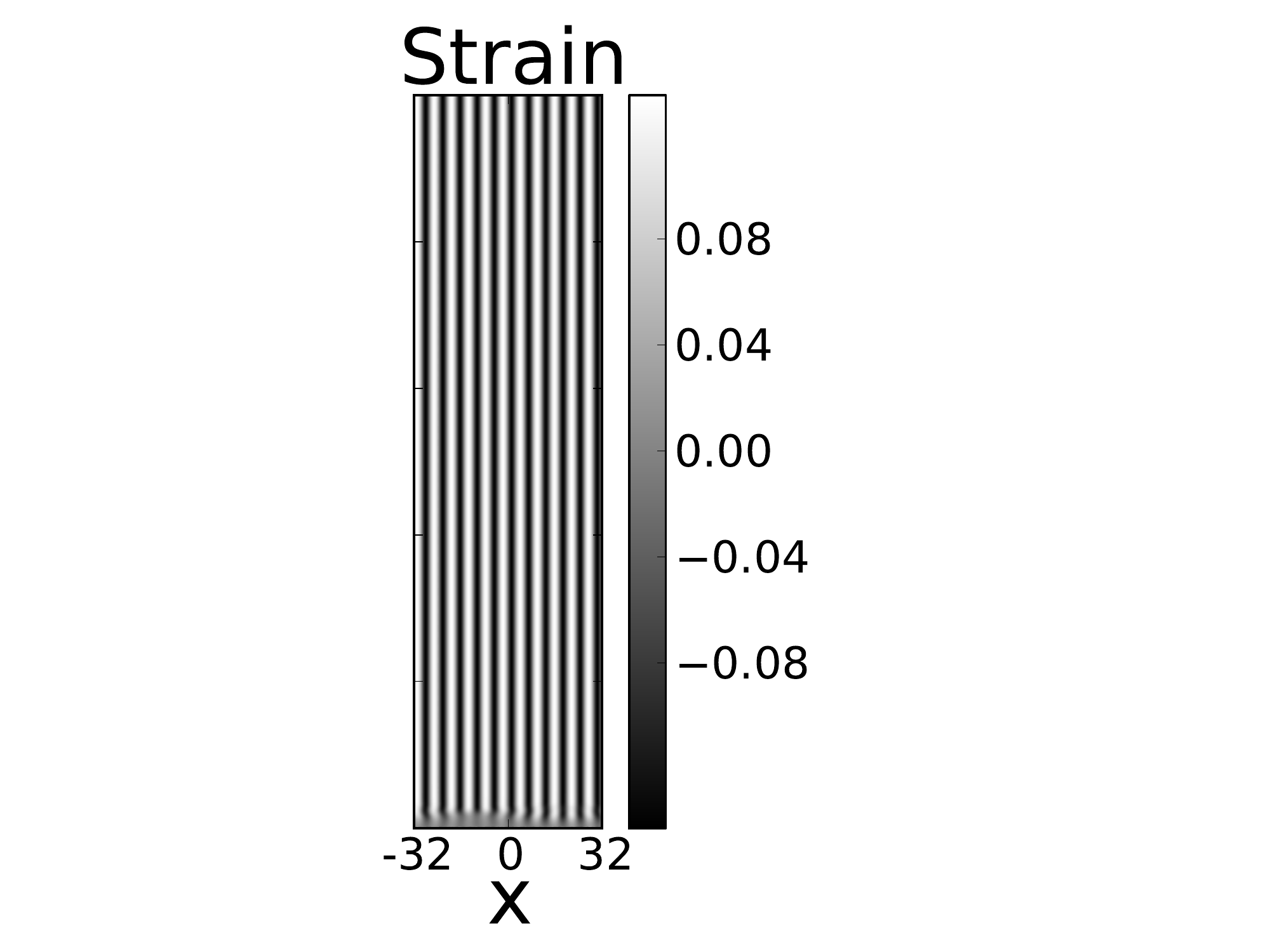}
\hspace*{-3.cm}
\includegraphics*[width=0.43\columnwidth]{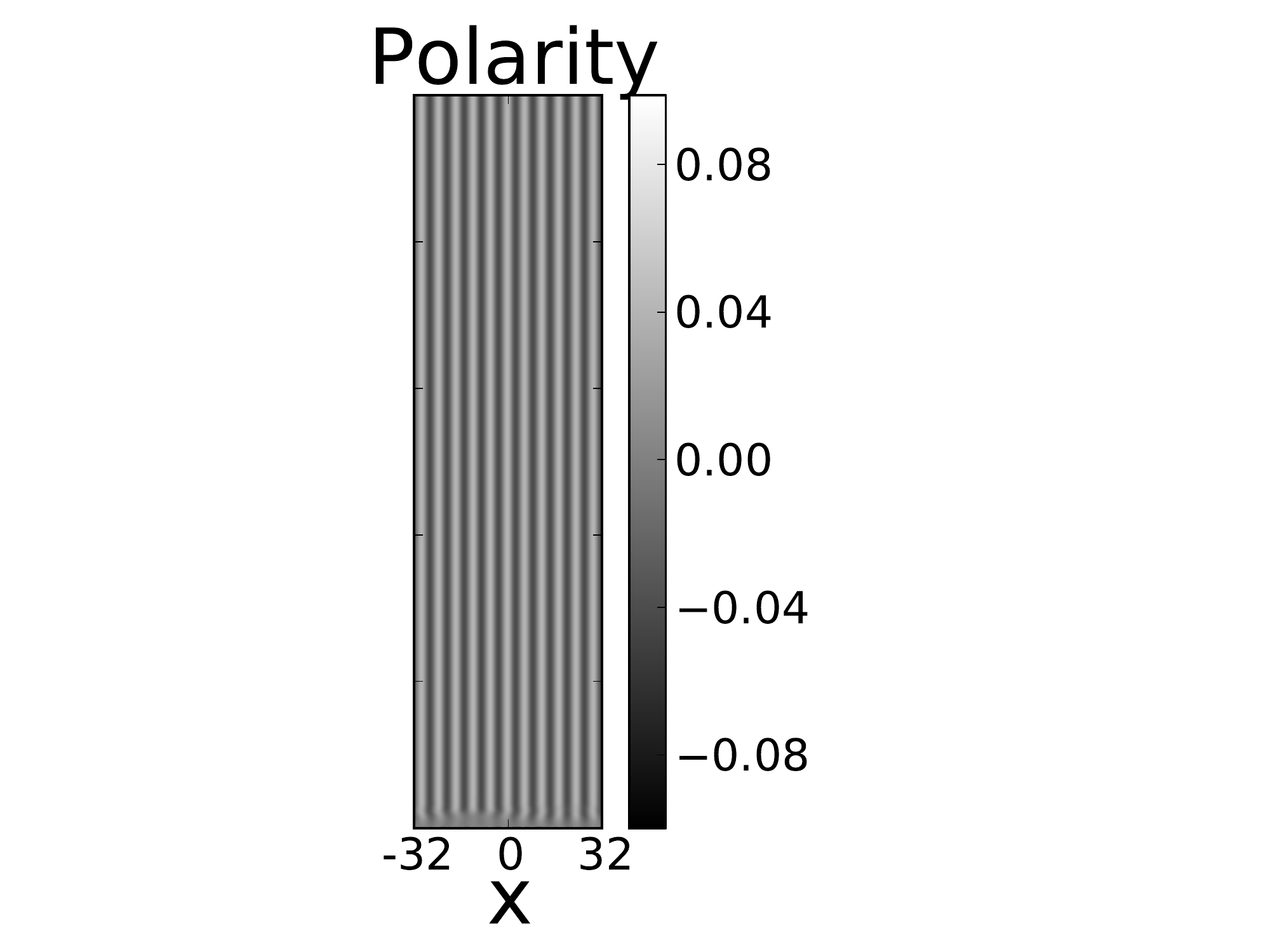}
\hspace*{-3.cm}\textbf{(c)}
\caption{
\textbf{Stationary bifurcation.}  
Numerical simulations of 
eqs.~(\ref{eq:dynsig}-\ref{eq:dynp})
were performed with \texttt{XMDS}2 \cite{Dennis2013}, with
random initial conditions and periodic boundary conditions.
The parameter values are: 
$a = 1$, $K = 1$, $w = 1$, 
$\tau = \tau_{\mathrm{e}} = \tau_{\mathrm{p}} = 1$, 
$v_{\mathrm{p}} = 0$, 
$\lambda = 1$, 
$\alpha_{\mathrm{a}} = 2$, 
$\Gamma_{\mathrm{p}} = 1$, 
$\nu = 1$,
leading to a threshold $l_{\mathrm{p}}^c = 6$, see eq.~(\ref{eq:condlp}).
Space-time plots of the velocity, strain and polarity fields are given for
$l_{\mathrm{p}} = 5.99$ (\textbf{a});   
$l_{\mathrm{p}} = 6.01$ (\textbf{b});
$l_{\mathrm{p}} = 6.1$ (\textbf{c}). 
}
\label{fig:maxwell:statbif} 
\end{figure}

\section{Instabilities}
\label{sec:stab}

\subsection{Stationary bifurcation}
\label{sec:stab:stat}

Harmonic perturbations of amplitude $(v_0, e_0, p_0)$ to the 
reference homogeneous state read 
$(v, e, p) = (v_0, e_0, p_0) \; e^{s(q) t - i q x }$, 
defining the growth rate $s(q)$ at wavenumber $q$.
To linear order in the amplitudes $(v_0, e_0, p_0)$, eq.~(\ref{eq:dyne})
allows to eliminate the velocity  ($v_0 =  v_{\mathrm{p}} p_0 +  \frac{i}{q} 
(s + \frac{1}{\tau_{\mathrm{e}}}) e_0$), and to obtain a system of two linear
equations in the unknowns $(e_0, p_0)$.
Cancelation of the determinant yields a second-order polynomial
equation $s^2 + B s + C = 0$ for the growth rate.
Since the coefficient 
$B(q^2)  = \frac{1}{\tau} + \frac{1}{\tau_{\mathrm{e}}}  
+ \frac{1}{\tau_{\mathrm{p}}}  +  
\Gamma_{\mathrm{p}} (  K + \nu q^2 ) \, q^2$ 
is always positive, complex solutions 
for the growth rate are always damped: linear stability analysis 
of the homogeneous state precludes oscillatory instabilities.

However, the coefficient
$C(q^2)= 
\left( \frac{1}{\tau} + \frac{1}{\tau_{\mathrm{e}}}  \right)
\frac{1}{\tau_{\mathrm{p}}} +
\Gamma_{\mathrm{p}} 
\left[
K \left( \frac{1}{\tau} + \frac{1}{\tau_{\mathrm{e}}}  \right)
- w \left( v_{\mathrm{p}} + \frac{l_{\mathrm{p}}}{\tau} \right) 
\right]
\, q^2
+ \Gamma_{\mathrm{p}} \nu 
\left( \frac{1}{\tau} + \frac{1}{\tau_{\mathrm{e}}}  \right) \, q^4
$
may change sign, and a stationary instability occurs 
close to wavenumbers such that  $C(q^2) < 0$, 
with a growth rate $s_+ = (-B + \sqrt{B^2 - 4 C})/2 > 0$.
A straightforward calculation gives the instability threshold:
\begin{equation}
  \label{eq:condlp}
w \left( l_{\mathrm{p}} + v_{\mathrm{p}} \tau \right) > 
\left( 1 + \frac{\tau}{\tau_{\mathrm{e}}}  \right)
 \left( K + 2 \sqrt{\nu a} \right)
\end{equation}
or equivalently
\begin{equation}
  \label{eq:condlp:2}
w \left( \beta_{\mathrm{a}} + \eta \psi_{\mathrm{a}} \right) >
( 1 + \Gamma_{\mathrm{e}} \eta ) \, ( K G - w^2 + 2 G \sqrt{\nu a} ).
\end{equation}
Since $\eta \ge 0$, $\Gamma_{\mathrm{e}} \ge 0$, $G \ge 0$
and $w^2 \le K  G$, the r.h.s. of eq.~(\ref{eq:condlp:2}) is positive:
the bifurcation is made possible by the presence of active, polar couplings.
The fastest growing wavenumber is given by the solution $q_0$ of 
$\frac{\mathrm{d} C}{\mathrm{d} q^2}_{|q^2 = q_0^2} = 0$, from which 
the wavelength $\lambda_0 = 2\pi/q_0$ observed immediately beyond threshold 
is calculated
\begin{equation}
  \label{eq:lambda}
  \lambda_0 = 
2 \pi \, \left[
\frac{1}{2 G \nu}
\left( w^2 - K G + w \, 
\frac{\beta_{\mathrm{a}} + \eta \psi_{\mathrm{a}}}{1 + \Gamma_{\mathrm{e}} \eta} 
\right)
\right]^{-\frac{1}{2}} \, .
\end{equation}
This expression reduces to eq.~(\ref{eq:kelvin:lambda}) in the limit 
$\Gamma_{\mathrm{e}} \to 0$, $\psi_{\mathrm{a}} \to 0$.

The results of the linear stability analysis are confirmed by 
numerical simulations \cite{Dennis2013} of the full nonlinear system
eqs.~(\ref{eq:dynsig}-\ref{eq:dynp}),
supplemented for simplicity by random initial conditions, and 
by periodic boundary conditions
(see figure~\ref{fig:maxwell:statbif}).
Although one expects a Maxwellian material
to be liquid at times long compared to the
viscoelastic time $\tau$, the strain field does not relax to zero,
and exhibits a stationary, spatially periodic pattern of non-zero 
amplitude, due to the bifurcation made possible by active, polar
couplings. Since the instability occurs in the bulk, these
observations should be relevant far from the boundaries 
independently of specific boundary conditions  \cite{Yoshinaga2010}.

\subsection{Propagating waves}
\label{sec:stab:wave}

The damped (or stabilized) Kuramoto-Sivashinsky equation captures 
the universal features of numerous pattern-forming systems 
\cite{Manneville1988,Misbah1994}, such as directional solidification 
or liquid column flow \cite{Manneville1988,Brunet2007}.
In one spatial dimension, it reads
\begin{equation}
  \label{eq:dampedKS}
  \partial_{t} \phi + \phi \, \partial_x \phi  
=  - \delta \, \phi - \partial_x^2 \phi  - \partial_x^4 \phi. 
\end{equation}
where $\phi(x,t)$ denotes the relevant field, and $\delta$ is
a positive damping coefficient.
Large wavenumbers, destabilized by negative diffusion,
are stabilized by the fourth-order spatial derivative. 
The advection term $\phi \, \partial_x \phi  $ is responsible 
for energy transfer from small to large wavenumbers.
With $\phi = \phi_0 \; e^{s(q) t - i q x }$, 
the growth rate $s(q)$ at wavenumber $q$ is $s(q) = - \delta + q^2 - q^4$:
a stationary bifurcation occurs for $\delta \le \delta_c = \frac{1}{4}$,
with a most unstable wavenumber $q_0^2 = \frac{1}{2}$. 
The aspect ratio is defined as the ratio between the system size 
and the wavelength $\lambda_0 = 2 \pi / q_0$. For large aspect ratio,
decreasing the (positive) damping coefficient  leads through
a cascade of bifurcations to a regime of spatio-temporal chaos: 
the Kuramoto-Sivashinsky equation is recovered in the limit $\delta \to 0$.
Secondary and higher-order instabilities of eq.~(\ref{eq:dampedKS})
were studied for small and intermediate aspect ratios in \cite{Misbah1994}
and \cite{Brunet2007} respectively.

\begin{figure}[!t]
\includegraphics*[width=0.43\columnwidth]{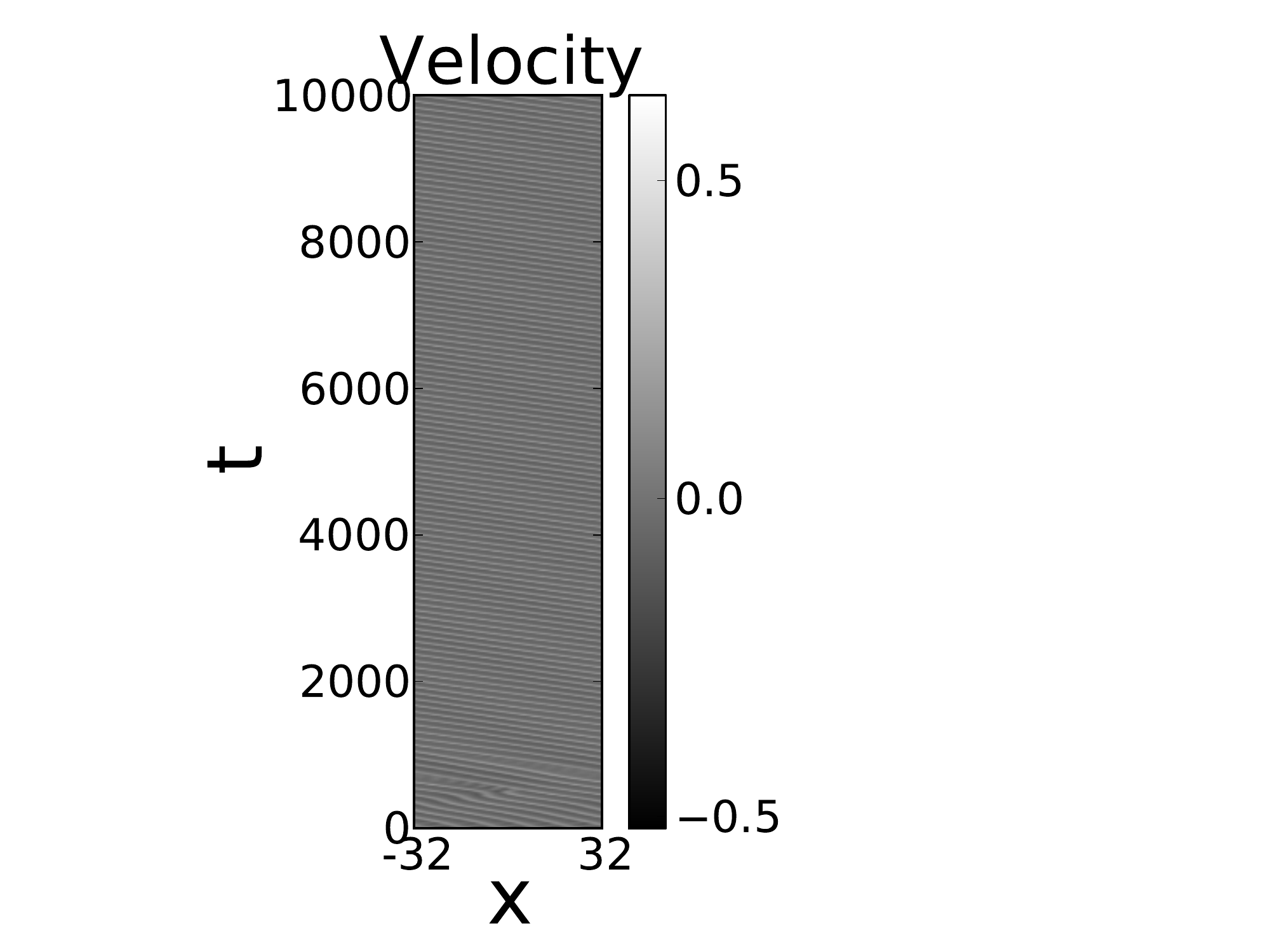}
\hspace*{-3.cm}
\includegraphics*[width=0.43\columnwidth]{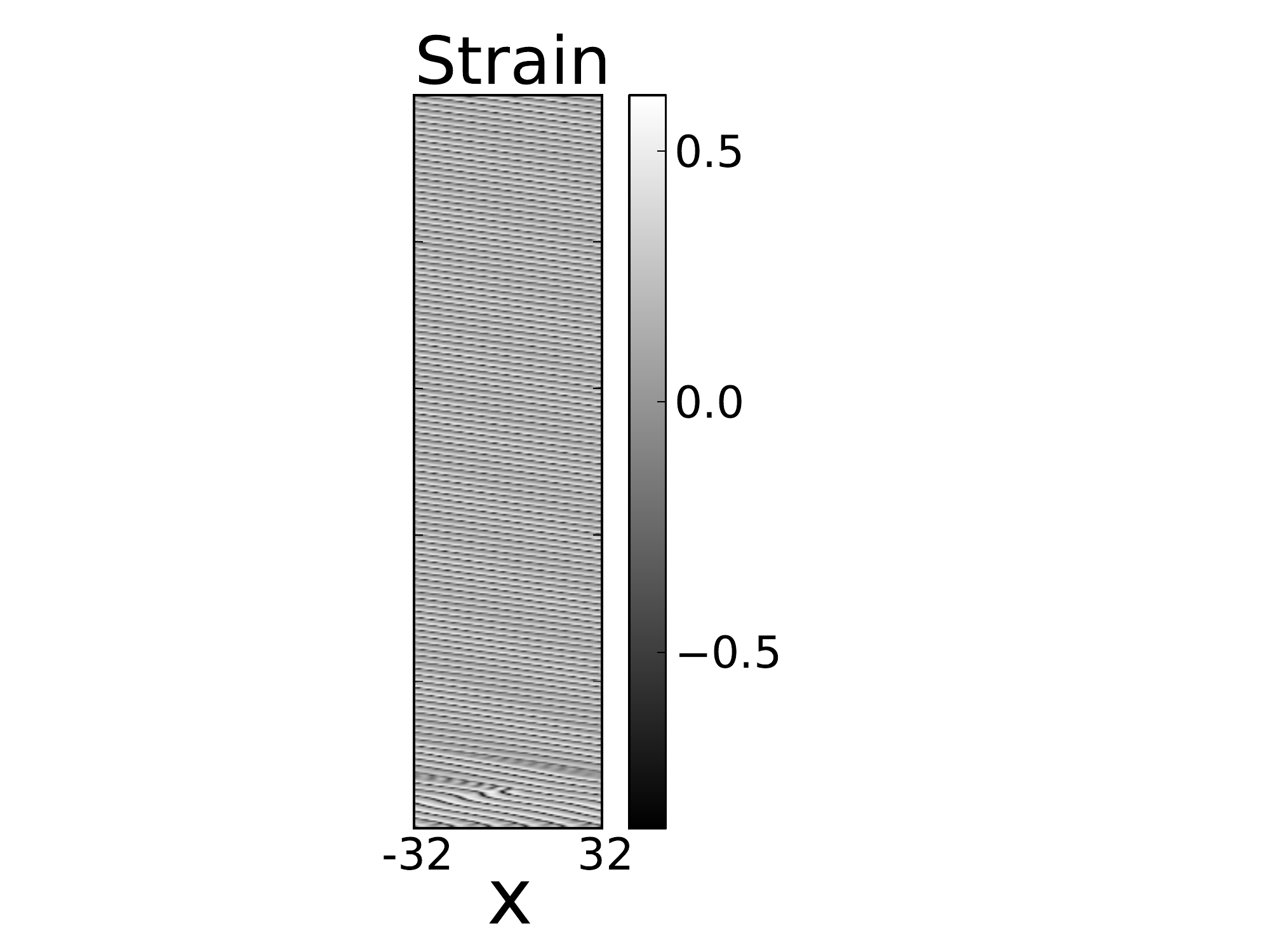}
\hspace*{-3.cm}
\includegraphics*[width=0.43\columnwidth]{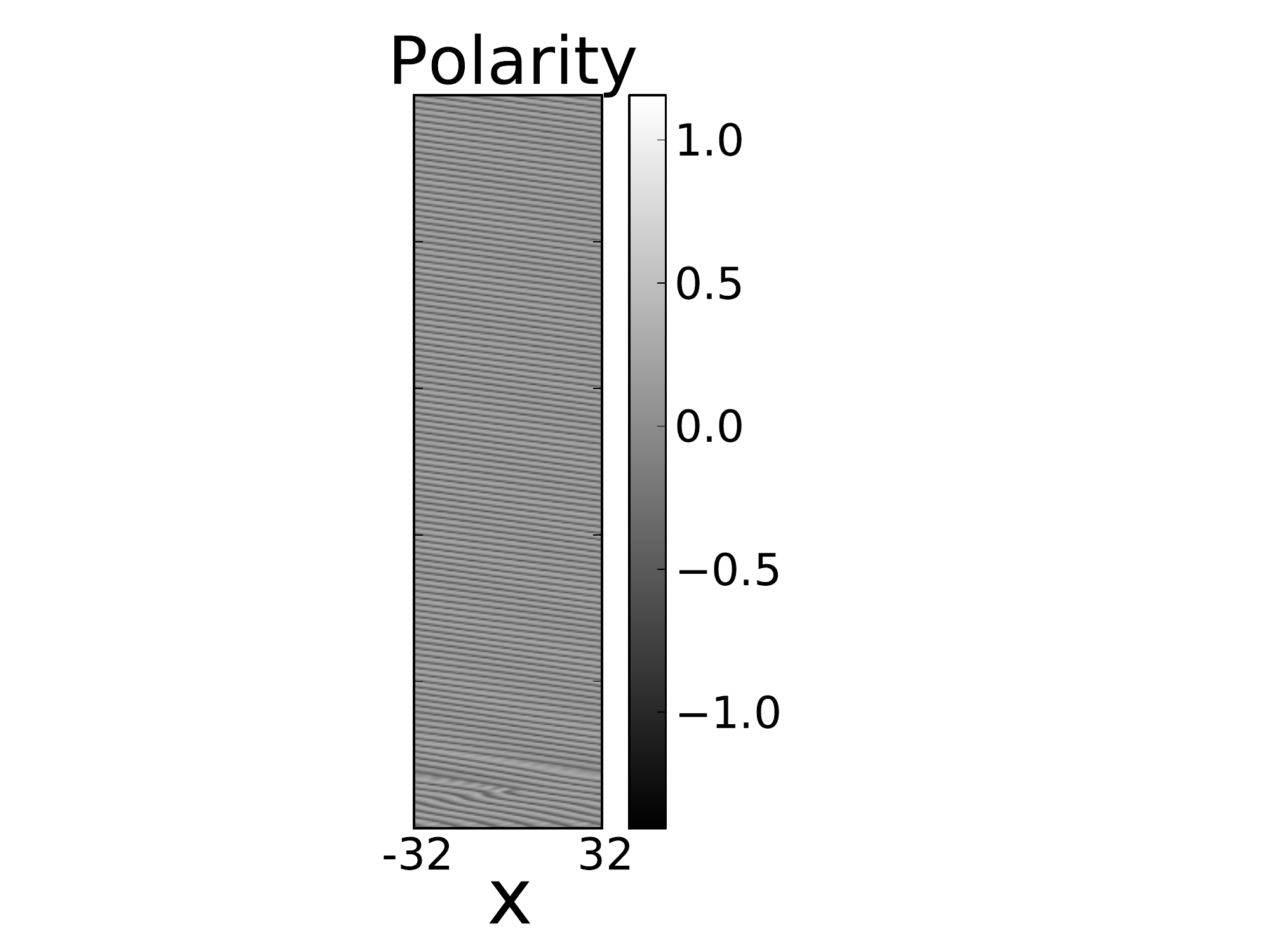}
\caption{
\textbf{Propagating waves.}
Numerical simulations of eqs.~(\ref{eq:dynsig}-\ref{eq:dynp})
were performed with \texttt{XMDS}2 \cite{Dennis2013}, with
periodic boundary conditions, and
periodic initial conditions, eqs.~(\ref{eq:init})
with $k = 0.5$ and $\varphi_0 = 0.5$.
The parameter values are: 
$a = 1$, $K = 1$, $w = 1$, 
$\tau = \tau_{\mathrm{e}} = 10$, 
$v_{\mathrm{p}} = 0$, 
$\lambda = 1$, 
$\alpha_{\mathrm{a}} = 2$, 
$\Gamma_{\mathrm{p}} = 0.1$ ($\tau_{\mathrm{p}} = 10$), 
$\nu = 1$,
Space-time plots of the velocity, strain and polarity fields are given for
$l_{\mathrm{p}} = 4.5$, larger than the threshold value 
$l_{\mathrm{p}}^c = 3.26$, see eq.~(\ref{eq:condlp}).
}
\label{fig:maxwell:propbif} 
\end{figure}

Since eq.(\ref{eq:dynp})
generalizes the damped Kuramoto-Sivashinsky equation by additional
couplings  with the velocity and strain fields, I conjecture that
their spatio-temporal dynamics contains that of the simpler system,
eq.~(\ref{eq:dampedKS}). 
Motivated by experiments (see section~\ref{sec:appl}), I use  
numerical simulations to confirm the existence of higher-order bifurcations 
of the homogeneous solution of 
eqs.~(\ref{eq:dynsig}-\ref{eq:dynp})
to propagating waves. 
Following \cite{Brunet2007}, appropriate initial conditions read
\begin{eqnarray}
  \label{eq:init:p}
  p(x,t = 0)  &=&  \sin(kx) + 0.5 \, \sin(2kx + \varphi_0) \nonumber \\ 
  \label{eq:init}
  e(x,t = 0)  &=&  0 \\
  \label{eq:init:v}
  v(x,t = 0) &=& -\frac{l_{\mathrm{p}}}{\tau} \, p(x,t = 0) \nonumber
\end{eqnarray}
where $k$ and $\varphi_0$ are a wavenumber and phase.
As shown in figure~\ref{fig:maxwell:propbif}, propagating waves are observed 
in numerical simulations of the system (\ref{eq:dynsig}-\ref{eq:dynp})
with initial conditions (\ref{eq:init}) and periodic
boundary conditions (see  figure~\ref{fig:kelvin:propbif}  
for propagating waves similarly obtained for an active, polar, 
viscoelastic solid). In this active, polar viscoelastic liquid, 
the strain field does not relax to zero for times large compared 
to the viscoelastic time $\tau$, but adopts the same patterns as 
the polarity and velocity fields to which it is coupled.

\section{Relevance to living matter}
\label{sec:appl}

The above analysis is based on the invariance properties of active 
polar materials. In sections~\ref{sec:appl:bundle} and \ref{sec:appl:tissue},
I respectively discuss possible
applications to the spatio-temporal dynamics 
of living matter obeying the same symmetries, 
first to the actin cytoskeleton, then to cohesive assemblies of collectively
migrating cells. In all cases, validating the model 
necessitates a detailed comparison with experimental data that is
left for future study.

\subsection{Actomyosin bundles}
\label{sec:appl:bundle}

A hydrodynamic description of the actin cytoskeleton requires 
length scales large compared to the mesh size of the polymer network. 
Actin filaments are polar: the hydrodynamic polarity field is naturally 
defined as a local, coarse-grained average of the polarity of individual 
actin filaments. Bundling is mediated by cross-linkers, either
active or passive, also responsible for most of the bundle's 
elasticity \cite{Yoshinaga2012}. 
A coupling between polarity and strain may arise given a
preference of cross-linkers to parallel or
anti-parallel pairs of actin filaments \cite{Meyer1990}. 
In the context of the actin cytoskeleton,
polarity advection is naturally interpreted as 
expressing the transport of polarity due to faster polymerization 
at the barbed ends of actin filaments.
Strain relaxes thanks to, \emph{e.g.}, crosslinker unbinding.

A stationary bifurcation is a likely scenario to explain the existence 
of monotonic (graded) and periodic (alternating) polarity patterns 
in actomyosin bundles \cite{Cramer1997},
that may self-organize into sarcomeres for large enough active couplings.
I show here that this bifurcation occurs whether the bundle rheology 
is that of a 
viscoelastic liquid or that of a viscoelastic solid, suggesting that the 
proposed mechanism does not depend on a specific rheology.
Assuming that myosins (resp. $\alpha$-actinins) concentrate close to 
the barbed (resp. pointed) ends of actin filaments, the same mechanism  
may also explain, with the same wavelength, 
the banded patterns seen in the distribution of myosin and
$\alpha$-actinin along stress fibers \cite{Peterson2004} and 
apical circumferential bundles \cite{Ebrahim2013}.
The wavelength of polarity patterns  depends
on the value of active couplings (see eqs.~(\ref{eq:lambda}) and 
(\ref{eq:kelvin:lambda})): a prediction of the model is that 
down- or up-regulating contractility should modify the typical 
length of a sarcomere. Indeed, this has been observed 
experimentally \cite{Peterson2004}.

Cells spread on a flat substrate assemble ring-like, 
contractile bundles around their nucleus \cite{Senju2009}.
Another prediction is that actomyosin rings may spontaneously rotate,
due to the occurrence of propagating waves 
in an active, polar, viscoelastic ring (see figures~\ref{fig:maxwell:propbif}  
and \ref{fig:kelvin:propbif}).
As a consequence of mechanical coupling between the actin cytoskeleton and 
the cell nucleus, rotation of the actin ring may in turn
entrain the rotation of the cell nucleus \cite{Kumar2014}. 
Another active, polar, viscoelastic ring is the contractile ring 
assembled during cytokinesis by eukaryotic cells \cite{Eggert2006}: it may 
also rotate thanks to the same mechanism. 

In \cite{Asano2009}, a propagating wave observed at the periphery of 
a spread, adhering cell under inhibition of Rac-mediated
lamellipodial formation was interpreted as an actin density wave
resulting from the active couplings of the polarity and concentration  
fields, in the limit of zero velocity in the actin cytoskeleton.
The model proposed here generalizes this result to a viscoelastic material
with arbitrary strain and velocity.

\subsection{Collective cell migration}
\label{sec:appl:tissue}

Hydrodynamic descriptions of tissues are possible for length scales 
large compared to a typical cell size. Motile cells are polar entities,
with a well-defined front and rear. A cohesive assembly
of cells migrating collectively \cite{Ilina2009,Roerth2012}
may for this reason be described as an active polar material \cite{Koepf2013}, 
where the coarse-grained polarity may arise from the rear-front polarity 
of individual cells \cite{Farooqui2005}, or from the position of 
the centrosome with respect to the cell nucleus \cite{Reffay2011}.
In \emph{Xenopus} mesendoderm cells, force application through 
cadherin-mediated adhesions leads to protrusive activity at the 
opposite side of the cell,
and to collective cell migration \cite{Weber2011}.
This coupling between cell-cell adhesion and polarity 
(see also \cite{desai2009}) leads to a possible interpretation of 
the coefficient $w$. A recent study relates strain relaxation
in tissues to cell division and apoptosis \cite{Ranft2010}.

Collective migration has been studied \emph{in vitro} in epithelial cell 
monolayers at confluence, with a recent report of 
propagating mechanical waves during tissue expansion \cite{Serra-Picamal2012}.
Interestingly, the wave velocity is distinct from the spreading
velocity of the epithelium boundary, and the phenomenon is
suppressed when inhibiting contractility: these observations
are consistent with the mechanism proposed here for 
propagating mechanical waves  in an active, polar, viscoelastic medium.
Note that the model admits propagating waves for an aspect ratio as small 
as $2$ (not shown), 
as observed in the experiment, albeit for different (free) 
boundary conditions \cite{Serra-Picamal2012}.

Strikingly, global tissue rotation appears to be a generic feature of 
epithelial cells (see \cite{Roerth2012} for a review), perhaps calling for 
a simple physical explanation such as the one proposed here.
An \emph{in vivo} example is the rotation of the follicular epithelium 
of \emph{Drosophila} eggs, contributing to the elongation 
of the egg \cite{Bilder2012}.
\emph{In vitro}, cell monolayers were found to 
rotate spontaneously on ring-shaped micro-patterned surfaces \cite{Wan2011}.
Whether cultured on a flat substrate \cite{Brangwynne2000}, 
or in a three-dimensional gel \cite{Tanner2012}
cell assemblies engage in coherent angular motion
for cell numbers as small as two, in which case a hydrodynamic 
description, if applicable, would be likely to pertain to cytoskeletal rather 
than epithelial  mechanics.

\section{Conclusion}
 \label{sec:conc}

Due to instabilities of the homogeneous state, the space-time dynamics 
of an active, polar, viscoelastic liquid in one dimension of space
includes stationary and propagating waves. On a ring, the propagation 
of mechanical and polarity waves is tantamount to spontaneous rotation.
Beyond the threshold of the linear instability, the strain field remains 
relevant for times long compared to the viscoelastic time:
the strain amplitude no longer relaxes to zero.
Necessary ingredients for pattern formation are a polarity advection term
$p \, \frac{\partial p}{\partial x}$,
as well as non-zero coupling terms between polarity gradients and the 
mechanical  fields.
The advection term arises due to symmetry
in hydrodynamic models of self-propelled systems \cite{Toner2005}, 
and has been explicitly derived from a microscopic model of the interaction of
actin filaments with molecular motors \cite{Ahmadi2006}.
Denoting $w$, $\beta_{\mathrm{a}}$ and  $\psi_{\mathrm{a}}$, 
the coupling coefficients between the polarity gradient and the strain,
stress, and strain rate field respectively, 
a necessary condition for the instability 
is that the combination $w (\beta_{\mathrm{a}} + \eta \psi_{\mathrm{a}})$  
be larger than a positive threshold value, where $\eta$ is the 
viscosity coefficient.

Since the evolution equation for the polarity field generalizes the
damped Kuramoto-Sivashinsky equation by including additional
couplings to the velocity and strain fields, it is tempting to
conjecture that the various dynamical states characteristic of the
damped Kuramoto-Sivashinsky equation are also relevant to active polar 
viscoelastic matter. At the scale of the cytoskeleton, predictions of 
the model may be best amenable to test \emph{in vitro},
using reconstituted bundles \cite{Claessens2006,Strehle2011,Thoresen2011},
or minimal systems composed of actin filaments, myosin minifilaments
and passive cross-linkers such as fascin or $\alpha$-actinin
\cite{Koehler2011}. Beyond stationary and propagating waves, one would 
like to know whether other secondary bifurcations \cite{Misbah1994,Brunet2007}, 
and perhaps space-time chaos, are relevant to cytoskeletal mechanics. 
A note of caution is however in order since additional (system-dependent)
nonlinearities may modify the present picture far from threshold.

A straightforward extension of this work is to include the strain-polarity 
couplings in constitutive equations for an active polar viscoelastic body
in the plane. Another natural testing ground may be \emph{in vitro} 
cohesive assemblies of collectively migrating cells. Indeed a 
disordered, large-scale cellular structure has been observed for the 
(hydrodynamic) velocity field, whose correlation
function decays exponentially \cite{Petitjean2010,Angelini2010}.
This behaviour is reminiscent of the cellular patterns 
of spatio-temporally chaotic regimes of the two-dimensional
Kuramoto-Sivashinsky equation \cite{Paniconi1997}.
Note that over long time scales, the density field in a tissue also depends 
on the rates of cell division and apoptosis \cite{Ranft2010}, 
an effect neglected here.

\section*{Acknowledgements}

I am grateful to C.~Blanch-Mercader, C.~Gay, F.~Graner, E.~Hannezo,
 J.~Prost, S.~Titli, B.~Vianay and N.~Yoshinaga 
for useful discussions, and would like to express special thanks
to J.-F.~Joanny, who motivated this study by mentioning one 
observation of a rotating contractile ring.

\section*{Appendix.  An active, polar, viscoelastic solid}
\appendix

To make this article self-contained, I outline the derivation of 
constitutive equations for an active, polar,
viscoelastic solid in one spatial dimension \cite{Yoshinaga2010}. 
Eq.~(\ref{eq:R}) becomes 
\begin{equation}
  \label{eq:kelvin:R}
  R = \left( \sigma + P  - \sigma^{\mathrm{el}}\right) \; \partial_x v 
      + \dot{p} \; h + r \; \Delta \mu
\end{equation}
leading to the constitutive equations 
 \begin{eqnarray}
  \label{eq:kelvin:consteqsig}
  \sigma +P  - \sigma^{\mathrm{el}} &=&  \eta \; \partial_x v + \sigma_{\mathrm{a}}
           +  \beta_{\mathrm{a}} \, \partial_x p  \\
  \label{eq:kelvin:consteqp}
  \dot{p} &=& \Gamma_{\mathrm{p}} \; h - \alpha_{\mathrm{a}}  \; p \, \partial_x p 
- \lambda  \; p \, \partial_x v \,. 
\end{eqnarray}
Eqs.~(\ref{eq:freeenergy}-\ref{eq:def:sigma:el}),
(\ref{eq:dynsig}) and (\ref{eq:dynp}) are unchanged.

\begin{figure}[!t]
\includegraphics*[width=0.43\columnwidth]{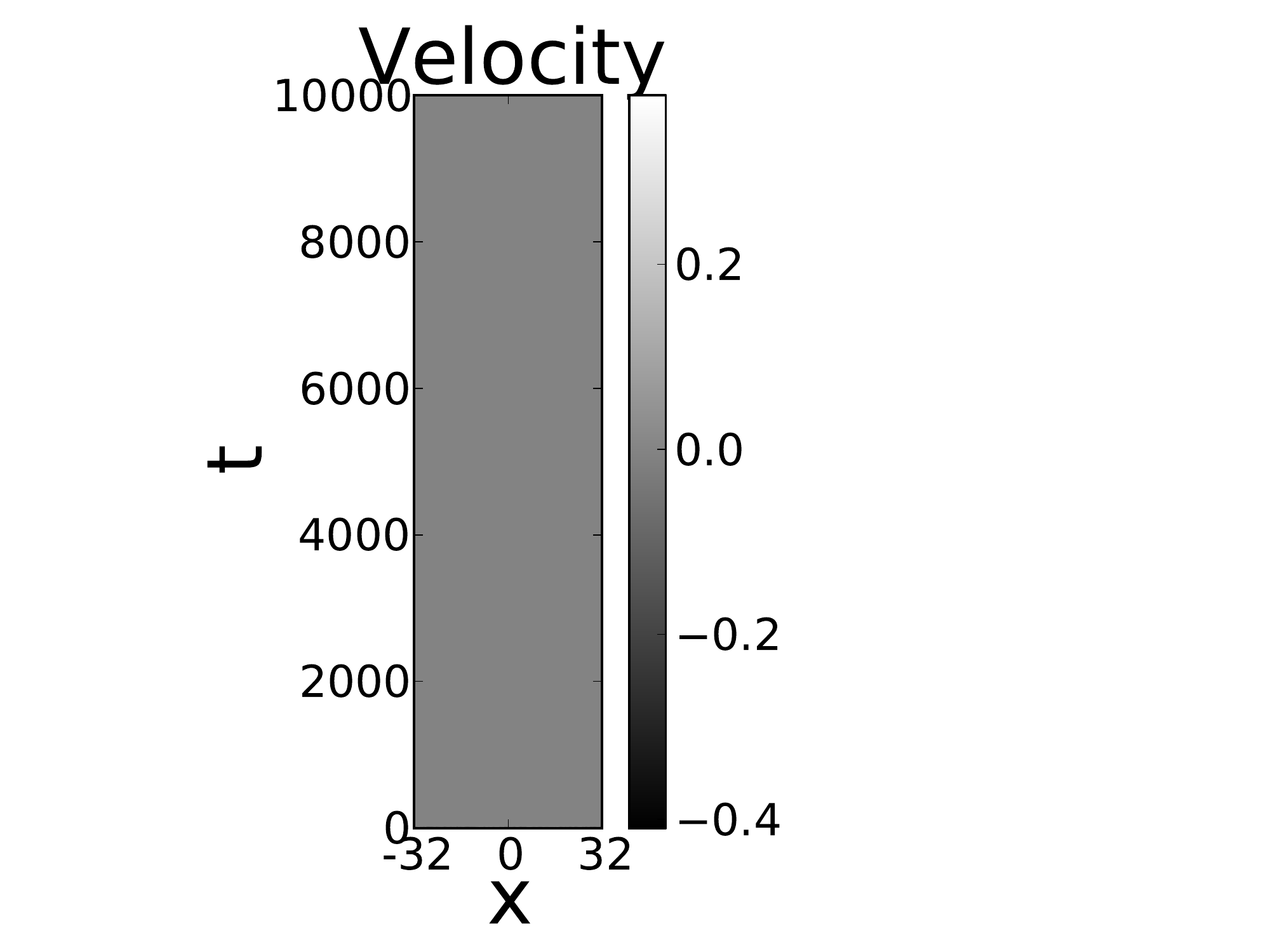}
\hspace*{-3.cm}
\includegraphics*[width=0.43\columnwidth]{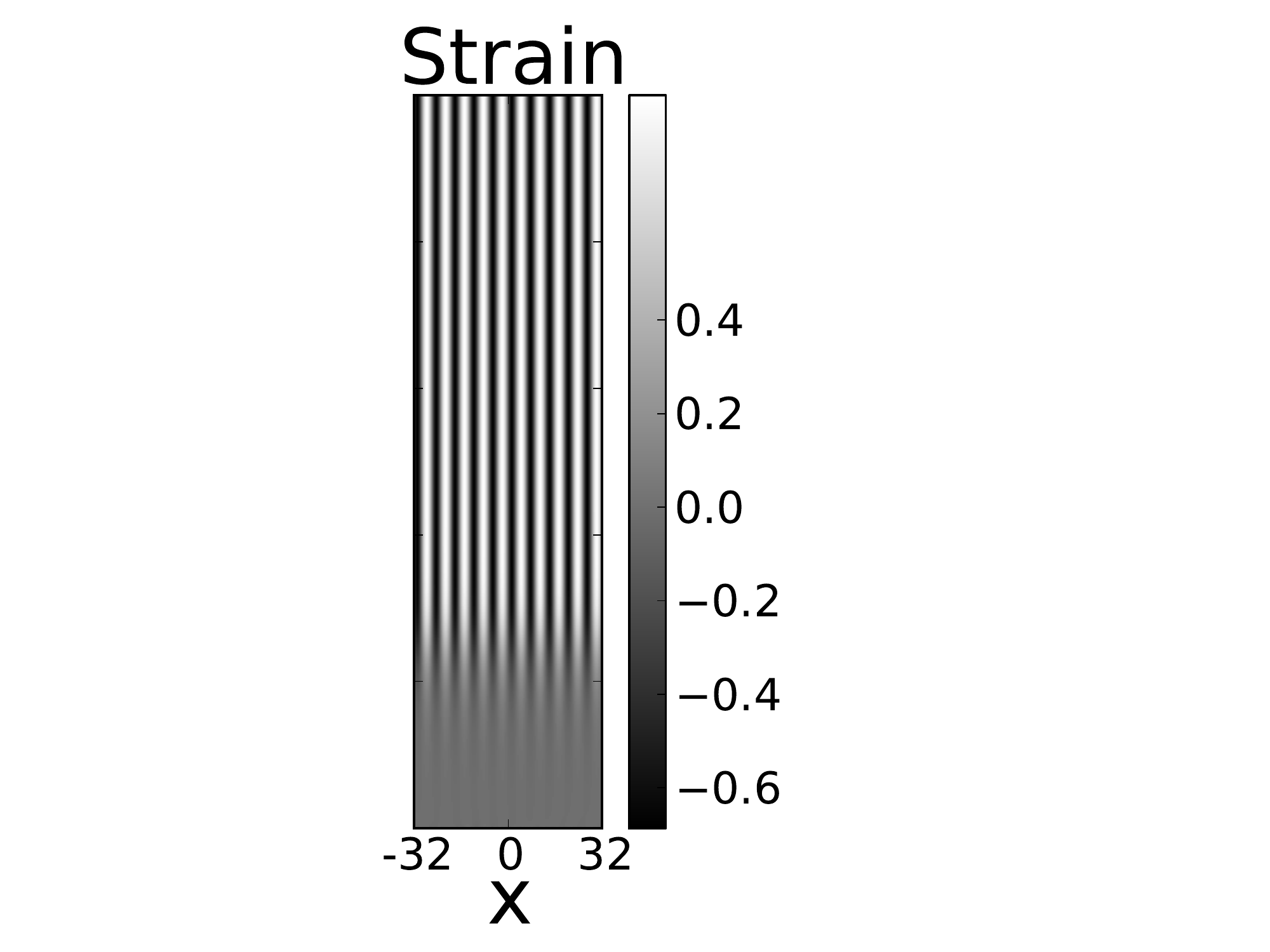}
\hspace*{-3.cm}
\includegraphics*[width=0.43\columnwidth]{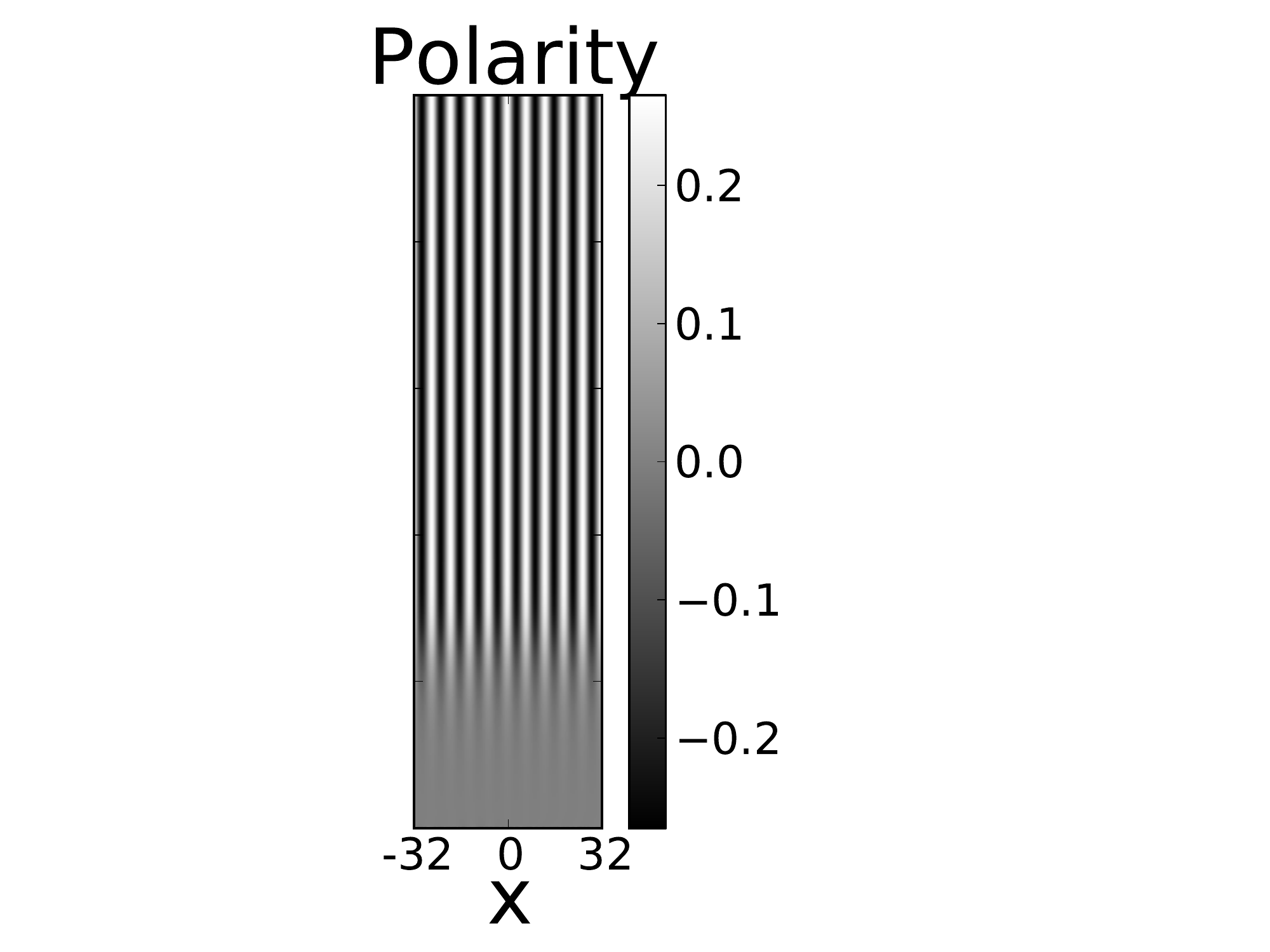}
\caption{
\textbf{Stationary bifurcation.}
Numerical simulations of 
eqs.~(\ref{eq:dynsig})-(\ref{eq:dynp})-(\ref{eq:kelvin:dyne}), 
were performed with \texttt{XMDS}2 \cite{Dennis2013}, with
random initial conditions and periodic boundary conditions.
The parameter values are: 
$a = 1$, $K = 1$, $w = 1$, 
$\tau =  \tau_{\mathrm{p}} = 1$, 
$\lambda = 1$, 
$\alpha_{\mathrm{a}} = 2$, 
$\Gamma_{\mathrm{p}} = 1$, 
$\nu = 1$,
leading to a threshold $l_{\mathrm{p}}^c = 3$, see eq.~(\ref{eq:kelvin:condlp}).
Space-time plots of the velocity, strain and polarity fields are 
shown immediately above threshold $l_{\mathrm{p}} = 3.01$.
}
\label{fig:kelvin:statbif} 
\end{figure}

In the particular case of a velocity field equal to zero, the strain field
is not transported (Eq.~(\ref{eq:dyne}) does not apply).
Upon substituting $\partial_x e$ from (\ref{eq:dynsig})
into (\ref{eq:dynp}), we found that the polarity field 
evolves according to the damped Kuramoto-Sivashinsky equation
\begin{equation}
  \label{eq:evol:kelvin:p}
  \partial_{t} p + \alpha_{\mathrm{a}}  \; p \, \partial_x p  
=  -\frac{1}{\tau_{\mathrm{p}}} \, p 
+ \Gamma_{\mathrm{p}} \left( K - w \, l_{\mathrm{p}} \right) \, \partial_x^2 p 
- \Gamma_{\mathrm{p}} \nu  \, \partial^4_x p \, .
\end{equation}
The polarity diffusion constant
$D_{\mathrm{p}} = \frac{\Gamma_{\mathrm{p}}}{G}\, 
(KG - w^2 - \beta_{\mathrm{a}}w )$
is positive in the absence of activity ($\beta_{\mathrm{a}} = 0$)
since $w^2 \le KG$. It may become negative in an active, polar material, 
leading to a stationary instability \cite{Yoshinaga2010}.

In the general case ($v \ne 0$), 
the strain field is transported according to 
$\dot{e} = \partial_x  v$, and eq.~(\ref{eq:dyne}) is replaced by
\begin{equation}
  \label{eq:kelvin:dyne}
\partial_t e + v \, \partial_x e   =  \partial_x v \,.     
\end{equation}
The linear stability analysis of the homogeneous state $v = e = p = 0$
unfolds as in section~\ref{sec:stab:stat}.
For the system (\ref{eq:dynsig})-(\ref{eq:dynp})-(\ref{eq:kelvin:dyne}), 
the necessary condition for a stationary bifurcation is
\begin{equation}
  \label{eq:kelvin:condlp}
w l_{\mathrm{p}} > 
K + 2 \sqrt{a \nu}
\end{equation}
also written as
\begin{equation}
  \label{eq:kelvin:condlp:2}
w \beta_{\mathrm{a}} >  KG - w^2 + 2 G \sqrt{a \nu}
\end{equation}
In agreement with \cite{Yoshinaga2010}, the wavelength of the stationary 
pattern is that of the most unstable mode 
\begin{equation}
  \label{eq:kelvin:lambda}
\lambda_0 = 2 \pi \sqrt{\frac{2 \nu}{w l_{\mathrm{p}} - K}}
= 2 \pi \sqrt{\frac{2 \nu G}{w \beta_{\mathrm{a}}  - (KG - w^2)}} \,.
\end{equation}
Depending on the experimental conditions, one may need to include
external friction in the force balance equation,
$\partial_x \sigma = \xi \, v$, with a positive friction coefficient $\xi$.
In the case of a viscoelastic solid, one can easily show that this
additional term leaves the threshold (\ref{eq:kelvin:condlp}) 
unchanged.

\begin{figure}[!t]
\includegraphics*[width=0.43\columnwidth]{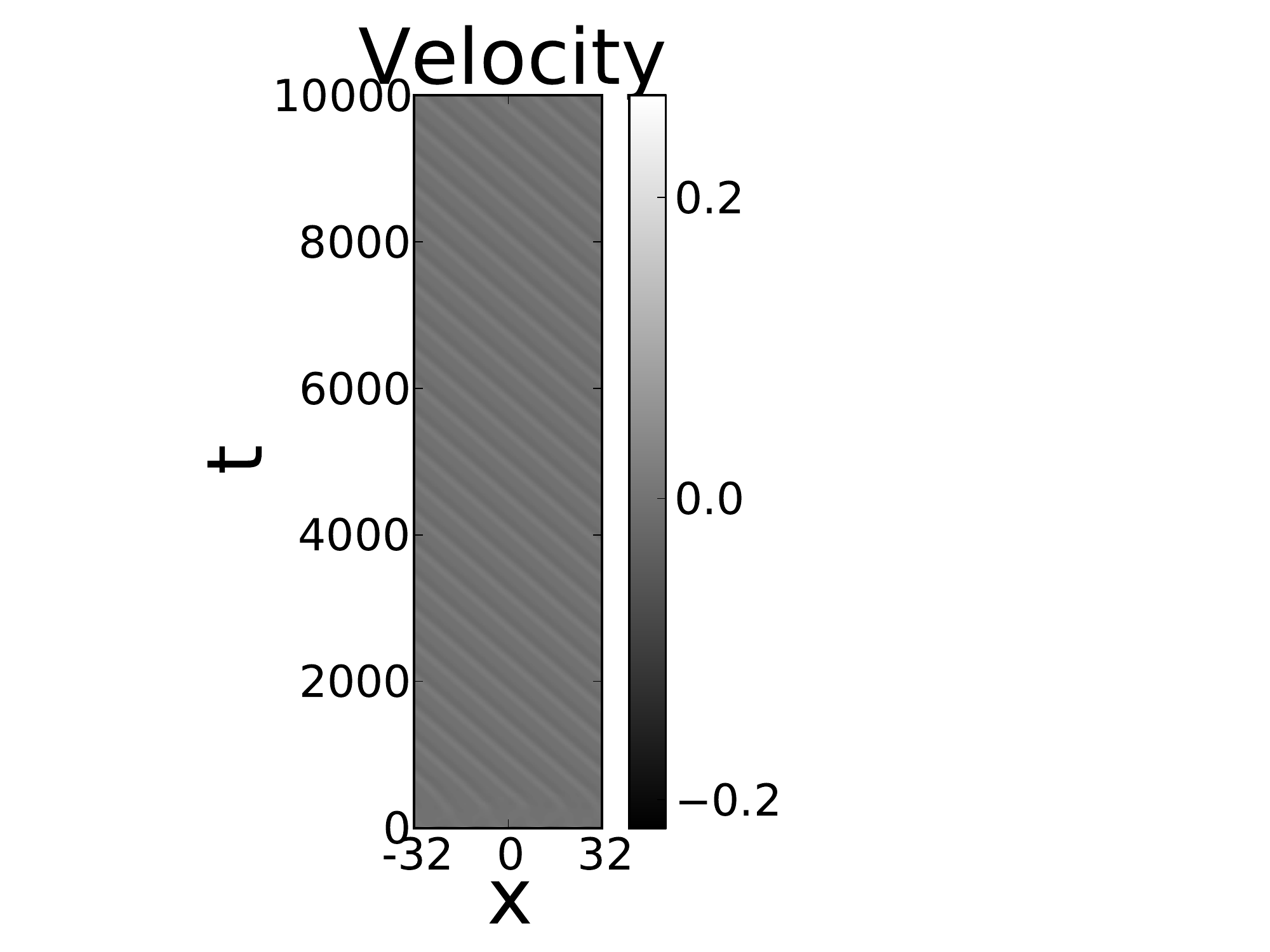}
\hspace*{-3.cm}
\includegraphics*[width=0.43\columnwidth]{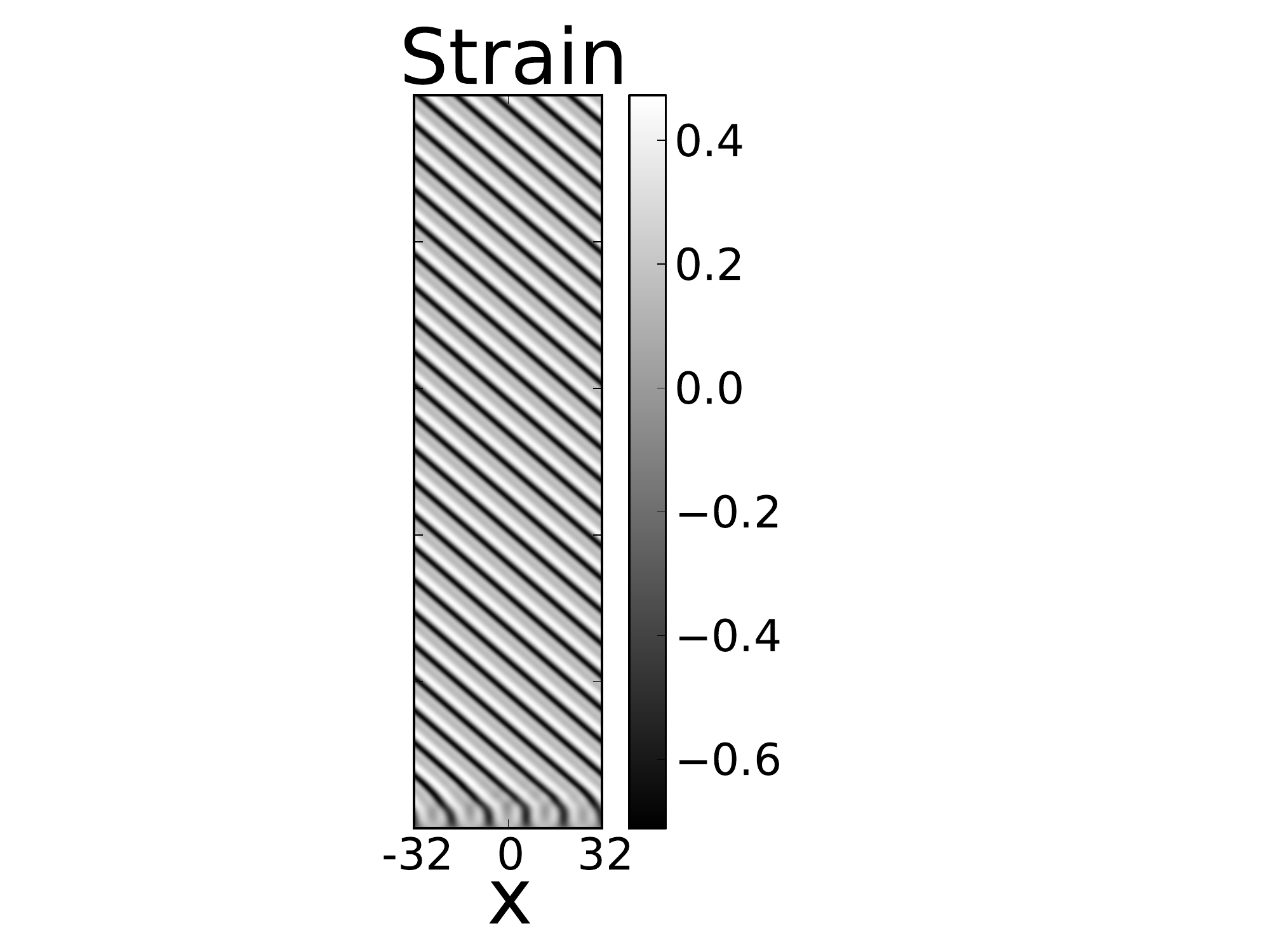}
\hspace*{-3.cm}
\includegraphics*[width=0.43\columnwidth]{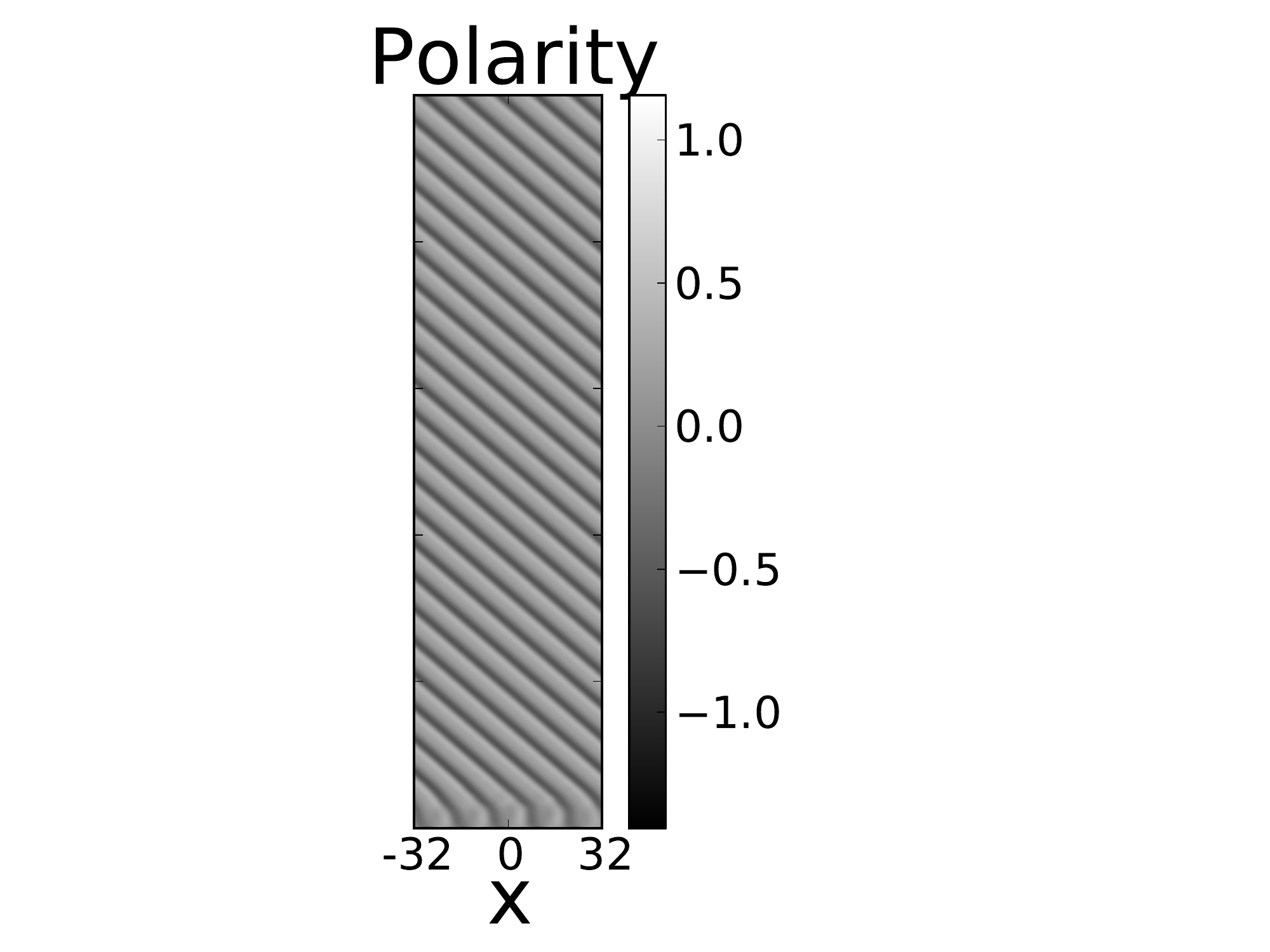}
\caption{
\textbf{Propagating waves.}
Numerical simulations of 
eqs.~(\ref{eq:dynsig})-(\ref{eq:dynp})-(\ref{eq:kelvin:dyne})
were performed with \texttt{XMDS}2 \cite{Dennis2013}, with
periodic initial conditions, eqs.~(\ref{eq:init}),
with $k = 0.5$ and $\varphi_0 = 0.5$,
and periodic boundary conditions.
Space-time plots of the velocity, strain and polarity fields 
are shown for the parameter values:  $a = 1$, $K = 1$, $w = 1$, 
$\tau = 10$,  $\lambda = 1$, 
$\nu = 1$, $\alpha_{\mathrm{a}} = 2$, $\Gamma_{\mathrm{p}} = 0.1$ 
($\tau_{\mathrm{p}} = 10$)  and  $l_{\mathrm{p}} = 1.9$,
above the threshold value $l_{\mathrm{p}}^c \simeq 1.63$
calculated from eq.~(\ref{eq:kelvin:condlp}).
}
\label{fig:kelvin:propbif} 
\end{figure}

Numerical simulations support the results of the linear stability analysis,
see figure~\ref{fig:kelvin:statbif}.
Above threshold, stationary periodic patterns arise for the strain and 
polarity fields, whereas the velocity field quickly relaxes to zero
(see eq.~(\ref{eq:kelvin:dyne})),
confirming the relevance of the analysis previously performed in
the case of a zero velocity field \cite{Yoshinaga2010}.
Using periodic initial conditions (\ref{eq:init}), numerical
simlulations yield propagating waves for the velocity, strain and
polarity fields at lower values of the polarity damping
(figure~\ref{fig:kelvin:propbif}).

\bibliographystyle{epj}
\bibliography{maxwell}

\begin{thebibliography}{49}

\bibitem{Juelicher2007}
F.~J\"ulicher, K.~Kruse, J.~Prost, J.F. Joanny, Phys. Rep. \textbf{449}, 3
  (2007)

\bibitem{Ramaswamy2010}
S.~Ramaswamy, Annu. Rev. Cond. Matt. Phys. \textbf{1}, 323 (2010)

\bibitem{Marchetti2013}
M.C. Marchetti et~al., Rev. Mod. Phys. \textbf{85}, 1143 (2013)

\bibitem{Chen2010}
D.~Chen et~al., Annu. Rev. Cond. Matter Phys. \textbf{1}, 301 (2010)

\bibitem{Kruse2004}
K.~Kruse et~al., Phys. Rev. Lett. \textbf{92}, 078101 (2004)

\bibitem{Kruse2005}
K.~Kruse et~al., Eur. Phys. J. E \textbf{16}, 5 (2005)

\bibitem{Muhuri2007}
S.~Muhuri et~al., Europhys. Lett. \textbf{78}, 48002 (2007)

\bibitem{Giomi2008}
L.~Giomi et~al., Phys. Rev. Lett. \textbf{101}, 198101 (2008)

\bibitem{Banerjee2011}
S.~Banerjee, T.B. Liverpool, M.C. Marchetti, Europhys. Lett. \textbf{96}, 58004
  (2011)

\bibitem{Edwards2011}
C.M. Edwards, U.S. Schwarz, Phys. Rev. Lett. \textbf{107}, 128101 (2012)

\bibitem{Yoshinaga2012}
N.~Yoshinaga, P.~Marcq, Phys. Biol. \textbf{9}, 046004 (2012)

\bibitem{Callan-Jones2011}
A.C. Callan-Jones, F.~J\"ulicher, New. J. Phys. \textbf{13}, 093027 (2011)

\bibitem{Yoshinaga2010}
N.~Yoshinaga, J.F. Joanny, J.~Prost, P.~Marcq, Phys. Rev. Lett. \textbf{105},
  238103 (2010)

\bibitem{Manneville1988}
P.~Manneville, \emph{The Kuramoto-Sivashinsky equation: a progress report}, in
  \emph{Propagation in systems far from equilibrium}, edited by J.E. Wesfreid
  et~al. (1988), Vol.~40 of \emph{Springer Series in Synergetics}, pp. 265--280

\bibitem{Brand1990}
H.~Brand, H.~Pleiner, W.~Renz, J. Phys. \textbf{51}, 1065 (1990)

\bibitem{Chaikin1995}
P.~Chaikin, T.~Lubensky, \emph{Principles of condensed matter physics}
  (Cambridge University Press, 1995)

\bibitem{Dennis2013}
G.R. Dennis, J.J. Hope, M.T. Johnsson, Comp. Phys. Comm. \textbf{184},
  201–208 (2013)

\bibitem{Misbah1994}
C.~Misbah, A.~Valance, Phys. Rev. E \textbf{49}, 166 (1994)

\bibitem{Brunet2007}
P.~Brunet, Phys. Rev. E \textbf{76}, 017204 (2007)

\bibitem{Meyer1990}
R.K. Meyer, U.~Aebi, J Cell Biol \textbf{110}, 2013 (1990)

\bibitem{Cramer1997}
L.P. Cramer et~al., J. Cell Biol. \textbf{136}, 1287 (1997)

\bibitem{Peterson2004}
L.J. Peterson et~al., Mol. Biol. Cell \textbf{15}, 3497 (2004)

\bibitem{Ebrahim2013}
S.~Ebrahim et~al., Curr. Biol. \textbf{23}, 731 (2013)

\bibitem{Senju2009}
Y.~Senju, H.~Miyata, J. Biochem. \textbf{145}, 137 (2009)

\bibitem{Kumar2014}
A.~Kumar et~al., Sci Rep \textbf{4}, 3781 (2014)

\bibitem{Eggert2006}
U.S. Eggert, T.J. Mitchison, C.M. Field, Annu. Rev. Biochem. \textbf{75}, 543
  (2006)

\bibitem{Asano2009}
Y.~Asano et~al., HFSP J. \textbf{3}, 194 (2009)

\bibitem{Ilina2009}
O.~Ilina, P.~Friedl, J. Cell Sci. \textbf{122}, 3203 (2009)

\bibitem{Roerth2012}
P.~R{\o}rth, EMBO Rep. \textbf{13}, 984 (2012)

\bibitem{Koepf2013}
M.H. K\"opf, L.M. Pismen, Soft Matter \textbf{9}, 3727 (2012)

\bibitem{Farooqui2005}
R.~Farooqui, G.~Fenteany, J Cell Sci \textbf{118}, 51 (2005)

\bibitem{Reffay2011}
M.~Reffay et~al., Biophys. J. \textbf{100}, 2566 (2011)

\bibitem{Weber2011}
G.F. Weber, M.A. Bjerke, D.W. Desimone, Dev. Cell \textbf{22}, 104–115 (2011)

\bibitem{desai2009}
R.A. Desai et~al., J. Cell Sci. \textbf{122}, 905 (2009)

\bibitem{Ranft2010}
J.~Ranft et~al., Proc. Natl. Acad. Sci. USA \textbf{107}, 20863 (2010)

\bibitem{Serra-Picamal2012}
X.~Serra-Picamal et~al., Nat. Phys. \textbf{8}, 628–634 (2012)

\bibitem{Bilder2012}
D.~Bilder, S.L. Haigo, Dev. Cell \textbf{22}, 12 (2012)

\bibitem{Wan2011}
L.Q. Wan et~al., Proc. Natl. Acad. Sci. USA \textbf{108}, 12295 (2011)

\bibitem{Brangwynne2000}
C.~Brangwynne et~al., In Vitro Cell Dev Biol Anim \textbf{36}, 563 (2000)

\bibitem{Tanner2012}
K.~Tanner et~al., Proc Natl Acad Sci USA \textbf{109}, 1973 (2012)

\bibitem{Toner2005}
J.~Toner, Y.~Tu, S.~Ramaswamy, Ann. Phys. \textbf{318}, 170 (2005)

\bibitem{Ahmadi2006}
A.~Ahmadi et~al., Phys. Rev. E \textbf{74}, 061913 (2006)

\bibitem{Claessens2006}
M.~Claessens, M.~Bathe, E.~Frey, A.R. Bausch, Nat. Mat. \textbf{5}, 748  (2006)

\bibitem{Strehle2011}
D.~Strehle et~al., Eur. Biophys. J. \textbf{40}, 93 (2011)

\bibitem{Thoresen2011}
T.~Thoresen, M.~Lenz, M.L. Gardel, Biophys. J. \textbf{100}, 2698 (2011)

\bibitem{Koehler2011}
S.~K\"ohler, V.~Schaller, A.R. Bausch, PLoS One \textbf{6}, e23798 (2011)

\bibitem{Petitjean2010}
L.~Petitjean et~al., Biophys. J. \textbf{98}, 1790 (2010)

\bibitem{Angelini2010}
T.E. Angelini et~al., Phys. Rev. Lett. \textbf{104}, 168104 (2010)

\bibitem{Paniconi1997}
M.~Paniconi, K.R. Elder, Phys. Rev. E \textbf{56}, 2713  (1997)

\end{thebibliography}

\end{document}